\documentclass[%
aip,
amsmath,amssymb,
reprint,%
]{revtex4-2}

\usepackage{graphicx}
\usepackage{dcolumn}
\usepackage{bm}

\usepackage{subfig}

\usepackage{epstopdf}
\usepackage{epsfig}

\usepackage[utf8]{inputenc}
\usepackage[T1]{fontenc}
\usepackage{mathptmx}
\usepackage{etoolbox}

\usepackage[colorlinks = true,
linkcolor = blue,
urlcolor  = blue,
citecolor = blue,
anchorcolor = blue]{hyperref}

\makeatletter
\def\@email#1#2{%
	\endgroup
	\patchcmd{\titleblock@produce}
	{\frontmatter@RRAPformat}
	{\frontmatter@RRAPformat{\produce@RRAP{*#1\href{mailto:#2}{#2}}}\frontmatter@RRAPformat}
	{}{}
}%
\makeatother
\begin{document}
	
	\preprint{AIP/123-QED}
	
	\title
	{Spectral and Dynamical Properties of the Fractional Nonlinear Schr\"odinger Equation under Harmonic Confinement}
	\author{R.\ Kusdiantara}
	\affiliation{Industrial and Financial Mathematics Research Group, Faculty of Mathematics and Natural Sciences,\\ Institut Teknologi Bandung,	Bandung, 40132, Indonesia}
	\affiliation{Center of Excellent in Predictive Risk and Simulation Modeling, Institut Teknologi Bandung, Bandung, 40132, Indonesia}
	
	\author{M.\ F.\ Adhari}%
	\affiliation{Industrial and Financial Mathematics Research Group, Faculty of Mathematics and Natural Sciences,\\ Institut Teknologi Bandung,	Bandung, 40132, Indonesia}
	
	\author{H.\ A.\ Mardi}
	\affiliation{Department of Mathematics, Universitas Syiah Kuala, Banda Aceh, 23111, Indonesia
	}%
	
	\author{I W.\ Sudiarta}
	\affiliation{Department of Physics, University of Mataram, Mataram, 83125, Indonesia
	}%
	
	\author{H.\ Susanto}
	\affiliation{Department of Mathematics, Khalifa University, PO Box 127788, Abu Dhabi, United Arab Emirates}
	\affiliation{Department of Mathematics, Faculty of Mathematics and Natural Sciences, Universitas Indonesia, Gedung D Lt.\ 2 FMIPA Kampus UI Depok, 16424, Indonesia}
	
	\email{hadi.susanto@yandex.com}
	
	\date{\today}
	
	\begin{abstract}
		We investigate the spectral and dynamical properties of the fractional nonlinear Schr\"odinger (fNLS) equation with harmonic confinement. In this setting, the classical Laplacian is replaced by its fractional power $(-\partial_x^2)^{\alpha/2}$ with $\alpha\in(1,2]$, introducing nonlocal, L\'evy-type dispersion. This modification fundamentally alters the balance between nonlinearity, dispersion, and trapping, reshaping both the structure and stability of stationary states. Using a Fourier pseudo-spectral discretization, we compute stationary branches as functions of the temporal frequency $\Omega$ in focusing ($\sigma=+1$) and defocusing ($\sigma=-1$) regimes, and assess spectral stability via the linearized eigenvalue problem. Direct simulations, performed with split-step and exponential time-differencing integrators, confirm these predictions and reveal $\alpha$-dependent transitions between coherent oscillations, bounded breathing dynamics, and decoherence or fragmentation. Our results show that decreasing $\alpha$ systematically shifts bifurcation curves, fragments stability windows for excited states, and amplifies instability in the focusing regime, while supporting robust coherence in the defocusing case. Beyond clarifying how harmonic confinement mediates the interplay between nonlinearity and fractional dispersion, the study also provides benchmarks for numerical treatments of fractional operators and points toward potential applications in nonlinear optics, Bose--Einstein condensates, and anomalous transport phenomena.
	\end{abstract}
	
	\maketitle
	
	
	\begin{quotation}
		The fractional nonlinear Schr\"odinger (fNLS) equation extends the classical model by using a fractional Laplacian that introduces nonlocal, Lévy-type dispersion. This feature allows the fNLS to describe anomalous transport and long-range interactions found in systems such as nonlinear optical media and Bose–Einstein condensates. When a harmonic trap is added, the balance between nonlinearity, dispersion, and confinement changes in a fundamental way. In this work, we study how the fractional order affects stationary states, bifurcation structures, and stability in both focusing and defocusing regimes. Using Fourier pseudo-spectral calculations and direct time simulations, we show that reducing the fractional order compresses and fragments stability windows, strengthens instability in focusing cases, and supports coherence in defocusing ones. These results explain how fractional dispersion and trapping interact and provide useful benchmarks for future studies.
	\end{quotation}

	\section{Introduction}\label{sec:intro}
	
	The fractional nonlinear Schr\"odinger (fNLS) equation generalizes the classical NLS by replacing the Laplacian with its fractional power $(-\partial_x^2)^{\alpha/2}$, $\alpha \in (1,2]$, thereby introducing nonlocal, L\'evy-type dispersion. This framework has attracted wide interest in applied mathematics, optics, and quantum physics, as fractional transport is a natural model for anomalous diffusion in complex or disordered media \cite{metzler2000random,bouchaud1990anomalous,west2003physics,shlesinger1995levy,tarasov2011fractional}. Laskin's work on fractional quantum mechanics established a systematic operator formulation \cite{laskin2002fractional}, while optical analogues demonstrated controllable platforms for fractional diffraction using photonic devices \cite{longhi2015fractional,malomed2021optical}. Mathematically, the fractional Laplacian is now a central tool in analysis and partial differential equations, with thorough expositions available on nonlocal operators and fractional Sobolev spaces \cite{diNezza2012hitchhiker}.
	
	Beyond theory, the fractional Laplacian has become increasingly relevant in experiments. In optics, graded-index and photonic lattice platforms enable tunable fractional diffraction, {providing engineered or effective realizations of fractional NLS-type models in controlled laboratory settings \cite{longhi2015fractional,malomed2021optical,iomin2021fractional}. Recent experiments have gone beyond linear propagation and demonstrated nonlinear fractional wave dynamics in optical systems, including the observation of fractional solitons governed by effective fractional nonlinear Schr\"odinger equations implemented via programmable dispersion engineering \cite{hoang2025nonlinear,hoang2026nyquist}.} In Bose-Einstein condensates, effective nonlocal dispersion can arise through long-range interactions or engineered optical lattices, linking the fNLS to condensed matter experiments \cite{stephanovich2024spin,stickler2013potential}. More broadly, fractional dynamics have been used to describe energy transport in disordered or fractal media \cite{metzler2000random,tarasov2011fractional}, turbulence cascades \cite{epps2018turbulence,zhong2025fractional}, and quantum control protocols based on L\'evy-flight statistics \cite{poboiko2025measurement}. These examples illustrate the interdisciplinary role of the fractional Laplacian, providing a unifying framework that connects rigorous mathematics, nonlinear wave theory, and experimental realizations in physics. {One of the first experimental demonstrations of effective fractional group-velocity dispersion (GVD) and associated nonlinear fractional wave dynamics in fiber lasers, modeled by generalized fractional Schr\"odinger-type equations in the temporal domain, was reported in \cite{liu2023experimental,liu2025experimental,hoang2025nonlinear,hoang2026nyquist}
	}. Davis \textit{et al.}~\cite{davis2001fractional} achieved fractional dispersion experimentally by encoding both amplitude and phase components of the fractional derivative operator onto a phase-only liquid-crystal spatial light modulator (LCSLM), which functioned as a programmable Fourier-plane filter. {More recently, signatures of fractional dispersion have been demonstrated in optical systems via specifically designed experimental configurations, such as passively mode-locked fiber lasers incorporating an intracavity pulse shaper~\cite{hoang2025nonlinear}. 
		The pulse shaper imparts a tailored spectral phase that compensates for higher-order dispersion in the optical fiber while imposing a controlled fractional phase profile, thereby producing an effective fractional group-velocity dispersion within the laser cavity. We emphasize that, in most current experimental settings, fractional dispersion does not arise as a fundamental property of the medium, but instead represents an effective or engineered description, implemented over finite scales through tailored optical or photonic structures}.
	
	Harmonic trapping plays an essential role in nonlinear wave physics, for instance, in Bose-Einstein condensates and nonlinear optics, where it models confinement by magnetic/optical fields or graded-index media. For the classical case $\alpha=2$, stationary states, bifurcations, and stability under a harmonic potential are well established \cite{kivshar2001nonlinear,zezyulin2008stability,bao2004computing,bao2013numerical,susanto2011josephson,susanto2012josephson}.{When $1<\alpha<2$, harmonic confinement reshapes both the linear spectrum and the nonlinear branches, so that ground states and excited states may change width, amplitude, and stability as $\alpha$ departs from the classical limit \cite{kirkpatrick2016fractional,qiu2020stabilization}. 
		These changes are qualitative as well as quantitative, because the Fourier multiplier $|k|^\alpha$ modifies the balance between trapping, dispersion, and nonlinearity.
		In the linear regime, the propagation dynamics of wave packets governed by the fractional Schr\"odinger equation with a harmonic potential were investigated by Zhang et al.~\cite{zhang2015propagation}, where anharmonic oscillatory motion and irregular, decoherence-like behavior were reported.
	}
	
	Two central questions follow. First, how do stationary branches and their stability depend on the fractional order $\alpha$ in the presence of a harmonic trap? Second, how do spectral predictions connect to nonlinear time dynamics, including coherent persistence, spreading, or decoherence when fractional dispersion is present? While earlier studies have addressed aspects of existence and qualitative behavior \cite{laskin2002fractional,longhi2015fractional,kirkpatrick2016fractional,qiu2020stabilization}, systematic comparisons across $\alpha$ that combine stationary analysis, spectral stability, and time evolution within one framework remain limited.
	
	In this paper, we investigate spectral and dynamical properties of the trapped fNLS. Using a Fourier pseudo-spectral discretization with an explicit operator form of the fractional Laplacian, we continue stationary branches versus frequency $\Omega$ in both focusing and defocusing regimes, compute spectral stability by solving the linearized eigenvalue problem, and connect these predictions to direct time simulations. Our results show that decreasing $\alpha$ shifts bifurcation curves, fragments stability windows for excited states, and produces $\alpha$-dependent transitions between coherent oscillations and decoherence or fragmentation. These findings provide practical benchmarks for fractional operators and clarify how harmonic trapping mediates the interplay between nonlinearity and nonlocal dispersion.
	
	The remainder of this paper is organized as follows. 
	Section~\ref{sec:model} introduces the mathematical formulation of the fractional NLS and the definition of the fractional Laplacian under harmonic confinement. 
	Section~\ref{sec:methods} describes the numerical framework, including the pseudo-spectral discretization and the stability analysis. 
	The bifurcation diagrams and representative stationary states are presented in Section~\ref{sec:stat}. 
	In Section~\ref{sec:dynamics}, we connect the spectral predictions to direct time dynamics and highlight the transitions between coherent behavior and decoherence. 
	Finally, Section~\ref{sec:concl} summarizes the main findings and outlines future research directions.
	
	\section{Mathematical Model}\label{sec:model}
	
	We study the one-dimensional fNLS equation with a harmonic potential,
	\begin{equation}
		i\,\partial_t \psi
		= \left(-\partial_x^{2}\right)^{\alpha/2}\psi
		+ x^{2}\psi
		- \sigma |\psi|^{2}\psi,
		\label{eq:fnls}
	\end{equation}
	where $\psi(x,t)$ is a complex-valued wave function, $x\in\mathbb{R}$, $t\ge 0$, and $\alpha\in(1,2]$ denotes the order of the fractional Laplacian. The parameter $\sigma=\pm1$ distinguishes between the focusing $(+1)$ and defocusing $(-1)$ cases \cite{sulem2007nonlinear}. In the present analysis, the regime $\alpha\in[0,1]$ is omitted, as it typically leads to unstable solutions prone to collapse. Previous studies indicate that in this range, the solutions can undergo blow-up within finite evolution time \cite{duo2021dynamics,klein2014numerical,chen2018optical,boulenger2016blowup,klein2025numerical}.
	{Throughout the paper, figures that display both focusing and defocusing results (in particular Figs.~\ref{fig:bifur_alpha} and \ref{fig:spectrum_Re_alpha=2}--\ref{fig:alpha_vs_omega}) combine two independently computed solution families. The focusing ($\sigma=+1$) and defocusing ($\sigma=-1$) cases correspond to different equations and are treated separately. No continuation is performed across $\sigma$; the two cases are shown together only for comparison.
	}
	
	The fractional Laplacian $\left(-\partial_x^{2}\right)^{\alpha/2}$ is most naturally defined in Fourier space by the multiplier $|k|^{\alpha}$. To make this definition rigorous, we first consider functions $\psi$ in the Schwartz space $\mathcal{S}(\mathbb{R})$. This space consists of smooth functions on $\mathbb{R}$ that, together with all their derivatives, decay faster than any power of $|x|$ as $|x|\to\infty$. In other words, elements of $\mathcal{S}(\mathbb{R})$ are infinitely differentiable and vanish rapidly at infinity, which ensures that both $\psi$ and its Fourier transform are well behaved. Within this setting, we adopt the Fourier transform
	\begin{equation}
		(\mathcal{F}\psi)(k) := \hat{\psi}(k)
		= \frac{1}{2\pi}\int_{-\infty}^{\infty} e^{-ikx}\psi(x)\,dx,
	\end{equation}
	with inverse
	\begin{equation}
		(\mathcal{F}^{-1}\hat{\psi})(x)
		= \int_{-\infty}^{\infty} e^{ikx}\hat{\psi}(k)\,dk.
	\end{equation}
	
	In this convention,
	\begin{equation}
		\begin{array}{rcl}
			\displaystyle\left(-\partial_x^{2}\right)^{\alpha/2}\psi(x)
			&=&\displaystyle \int_{-\infty}^{\infty} |k|^{\alpha} e^{ikx}\hat{\psi}(k)\,dk,\\
			\displaystyle\mathcal{F}\left[\left(-\partial_x^{2}\right)^{\alpha/2}\psi\right](k) &=& \displaystyle|k|^{\alpha}\hat\psi(k).
		\end{array}
	\end{equation}
	This operator reduces to the standard Laplacian when $\alpha=2$, but for $1<\alpha<2$ it models anomalous diffusion and nonlocal interactions \cite{laskin2002fractional}.
	
	Equation~\eqref{eq:fnls} possesses the standard conserved quantities of the nonlinear Schrödinger type: the mass
	\begin{equation}
		Q = \int_{\mathbb{R}} |\psi(x,t)|^2\,dx,
		\label{eq:Q_norm}
	\end{equation}
	and the Hamiltonian energy
	\begin{equation}
		E = \int_{\mathbb{R}}
		\left(\tfrac{1}{2}\,\psi^* \left(-\partial_x^2\right)^{\alpha/2}\psi
		+ \tfrac{1}{2}\,x^2|\psi|^2
		- \tfrac{\sigma}{2}|\psi|^4 \right)dx,
	\end{equation}
	which remain constant for sufficiently regular solutions.
	
	Stationary, or standing-wave, solutions are sought in the form
	\[
	\psi(x,t) = e^{-i\Omega t}\,\tilde{\Phi}(x),
	\]
	where $\Omega$ is the frequency and $\tilde{\Phi}(x)$ is a real-valued profile. Substituting into \eqref{eq:fnls} yields the nonlinear eigenvalue problem
	\begin{equation}
		\Omega \tilde{\Phi}
		= \left(-\partial_x^2\right)^{\alpha/2}\tilde{\Phi}
		+ x^2 \tilde{\Phi}
		- \sigma \tilde{\Phi}^{3}.
		\label{eq:stat}
	\end{equation}
	
	In the linear case ($\sigma=0$) with $\alpha=2$, the eigenmodes are the Hermite functions,
	\begin{equation}
		\omega_n = 1+2n,
		\quad
		\Psi_n(x) = \frac{H_n(x)}{\sqrt{2^{\,n}\,\sqrt{\pi}\,(n!)}}\,e^{-x^2/2},
		\quad n=0,1,\dots,
	\end{equation}
	where $H_n(x)$ are the physicists’ Hermite polynomials,
	\[
	H_n(x) = e^{x^2}\frac{d^n}{dx^n}\left(e^{-x^2}\right).
	\]
	For $1<\alpha<2$, the eigenfrequencies of the linear problem are modified and can be expressed in terms of the Beta function \cite{laskin2002fractional}.
	
	In the nonlinear setting, solution branches are continued with respect to $\Omega$ for each nodal index $n=0,1,2,\dots$, and the corresponding $Q(\Omega)$ curves are displayed. 
	
	\section{Numerical Methods}\label{sec:methods}
	\subsection{Pseudo-spectral discretization}
	To approximate the fractional Laplacian in \eqref{eq:fnls}, we employ a Fourier pseudo-spectral method. The infinite line $\mathbb{R}$ is truncated to a periodic interval $[-L,L]$, discretized with $N$ uniform grid points
	\[
	x_j = -L + j\,dx, 
	\qquad j=0,1,\dots,N-1,
	\]
	where $dx = 2L/N$ is the spatial step size. 
	The parameters $L$ and $N$ are chosen to be sufficiently large to avoid boundary effects and maintain consistency throughout the numerical calculations. A $2\pi$-periodic function $F(x)$ can be represented by the truncated Fourier expansion
	\begin{equation}
		F(x_j) = \sum_{k=-N/2+1}^{N/2} \alpha_k e^{ikx_j},
		\label{eq:fourierseries}
	\end{equation}
	with Fourier coefficients $\alpha_k$. After reindexing $k' = k + N/2$, this can be rewritten in shifted form,
	\begin{equation}
		F(x_j) = \sum_{k=1}^{N} \alpha_{k-N/2}(-1)^k e^{i k x_j},
		\label{eq:fouriershift}
	\end{equation}
	and the coefficients are recovered by the inverse relation
	\begin{equation}
		\alpha_{k-N/2} = \frac{1}{N} \sum_{j=1}^{N} F(x_j)(-1)^j e^{-i k x_j}.
		\label{eq:inverse}
	\end{equation}
	The fractional Laplacian acts in Fourier space by multiplying each mode with $|k'|^{\alpha}$:
	\begin{equation}
		\left(-\frac{\partial^2}{\partial x^2}\right)^{\alpha/2}F(x_j)
		= \sum_{k=1}^{N} \alpha_{k-N/2}\,|k'|^{\alpha}\,e^{i k x_j}.
		\label{eq:fraclap}
	\end{equation}
	Equivalently, the action can be written in matrix form,
	\begin{equation}
		\left(-\frac{\partial^2}{\partial x^2}\right)^{\alpha/2} 
		= [a_{jk}][b_{jk}], \qquad j,k=1,\dots,N,
	\end{equation}
	with entries
	\begin{equation}
		a_{jk} = (-1)^j \big(k-N/2\big)^{\alpha} e^{i k x_j}, 
		\qquad
		b_{jk} = \frac{(-1)^k e^{-i j k}}{N}.
		\label{eq:psmatrix}
	\end{equation}
	
	In practice, the implementation is performed in \textsc{Matlab}. Instead of using the built-in \texttt{fft} routine, we adapt the Fourier differentiation matrix operator \texttt{FDMx} introduced by Shen and Tang \cite{tang2006spectral}. The key advantage of \texttt{FDMx} is that it provides an explicit operator form of the fractional Laplacian. This is essential for constructing the Jacobian in Newton’s method and for the spectral stability analysis of stationary states. By contrast, \texttt{fft} only evaluates transforms and does not directly yield the matrix form needed for linearization. This approach combines spectral accuracy with computational efficiency. 

	\subsection{Linearization and stability}
	To study the spectral stability of a stationary state, we perturb the real-valued profile $\tilde{\Phi}(x)$ as
	\begin{equation}
		\psi(x,t) = e^{-i\Omega t}\Big(\tilde{\Phi}(x) + \varepsilon\big(\bar{\Psi}_r(x,t) + i\,\bar{\Psi}_i(x,t)\big)\Big),
	\end{equation}
	where $\varepsilon$ denotes a small perturbation parameter. Substituting this expression into \eqref{eq:fnls} and keeping only terms linear in $\varepsilon$ yields a coupled system for the perturbations.
	Assuming perturbations with exponential time dependence,
	\[
	\bar{\Psi}_r(x,t) = \hat{\Psi}_r(x)e^{\lambda t}, 
	\qquad 
	\bar{\Psi}_i(x,t) = \hat{\Psi}_i(x)e^{\lambda t},
	\]
	we obtain the linear eigenvalue problem
	\begin{equation}
		\lambda
		\begin{pmatrix}
			\hat{\Psi}_r \\ \hat{\Psi}_i
		\end{pmatrix}
		=
		\begin{pmatrix}
			0 & -\mathcal{L}_{-} \\
			\mathcal{L}_{+} & 0
		\end{pmatrix}
		\begin{pmatrix}
			\hat{\Psi}_r \\ \hat{\Psi}_i
		\end{pmatrix},
		\label{eq:linstab}
	\end{equation}
	with operators
	\begin{equation}
		\begin{array}{rcl}
			\mathcal{L}_{-} &=& \Omega - x^2 + \sigma\tilde{\Phi}^2 - \left(-\partial_x^2\right)^{\alpha/2},\\
			\mathcal{L}_{+} &=& \Omega - x^2 + 3\sigma\tilde{\Phi}^2 - \left(-\partial_x^2\right)^{\alpha/2}.
		\end{array}
		\label{eq:linops}
	\end{equation}
	This system can be written compactly as
	\begin{equation}
		J \mathbf{u} = \lambda \mathbf{u}, 
		\qquad 
		J = 
		\begin{pmatrix}
			0 & -\mathcal{L}_{-} \\
			\mathcal{L}_{+} & 0
		\end{pmatrix}, 
		\qquad 
		\mathbf{u} =
		\begin{pmatrix}
			\hat{\Psi}_r \\ \hat{\Psi}_i
		\end{pmatrix},
		\label{eq:jacobian}
	\end{equation}
	where $J$ is the Jacobian operator obtained from the linearization.

	For numerical computation, the operators $\mathcal{L}_{\pm}$ are discretized using the Fourier pseudo-spectral framework introduced earlier. 
	The spectral stability of a stationary state $\tilde{\Phi}$ is determined by the eigenvalues of the Jacobian $J$. The state is classified as stable if all eigenvalues have nonpositive real parts, which guarantees that perturbations decay or remain bounded in time. Conversely, the state is unstable if at least one eigenvalue has a positive real part, leading to exponential growth of perturbations. This eigenvalue-based criterion provides a systematic framework for analyzing how stability is affected by the fractional order $\alpha$, the nonlinearity parameter $\sigma$, and the harmonic confinement. In particular, decreasing $\alpha$ tends to shift stability windows and introduces qualitative changes in the dynamics compared to the classical Schr\"odinger case $\alpha=2$.
	
	\subsection{Numerical time integration and diagnostics}
	
	\begin{figure*}[tbhp!]
		\centering
		\subfloat[$\alpha=2$]{\includegraphics[scale=0.375]{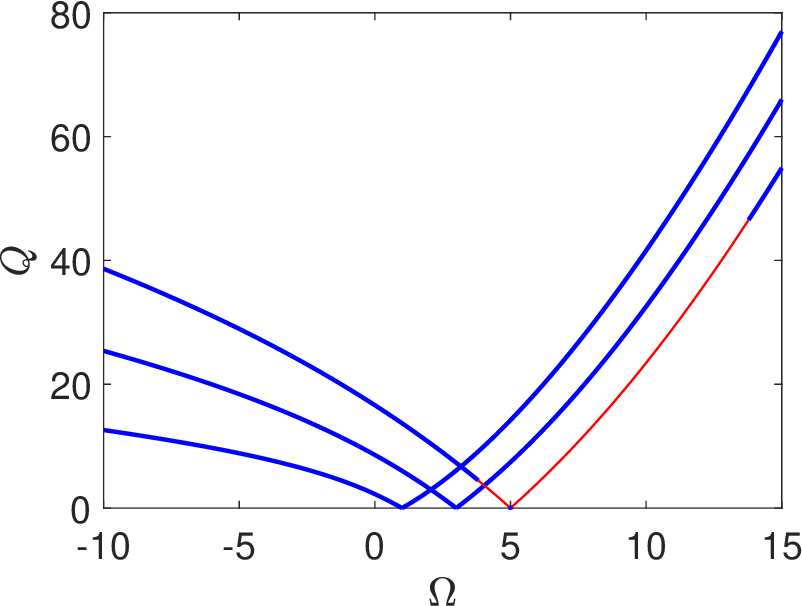}\label{subfig:bifur_alpha=2}}\quad
		\subfloat[$\alpha=1.5$]{\includegraphics[scale=0.375]{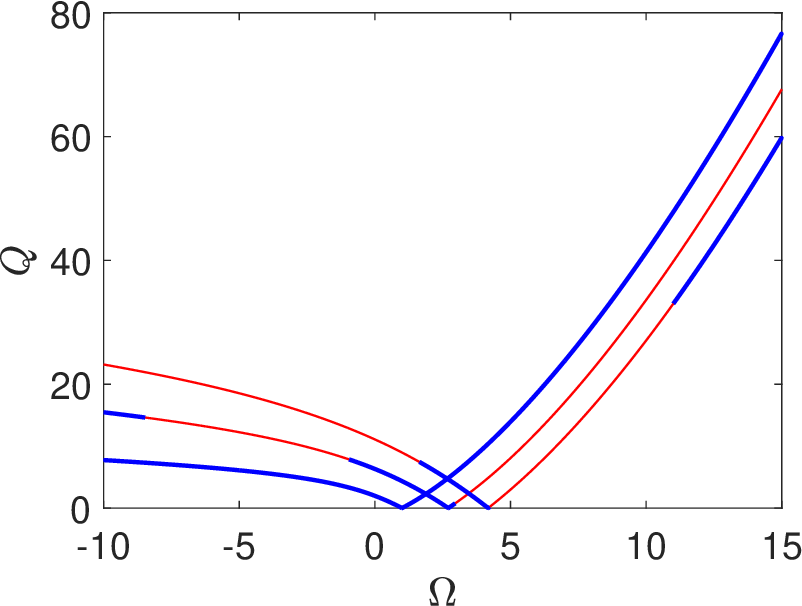}\label{subfig:bifur_alpha=1_5}}\quad
		\subfloat[$\alpha=1.1$]{\includegraphics[scale=0.375]{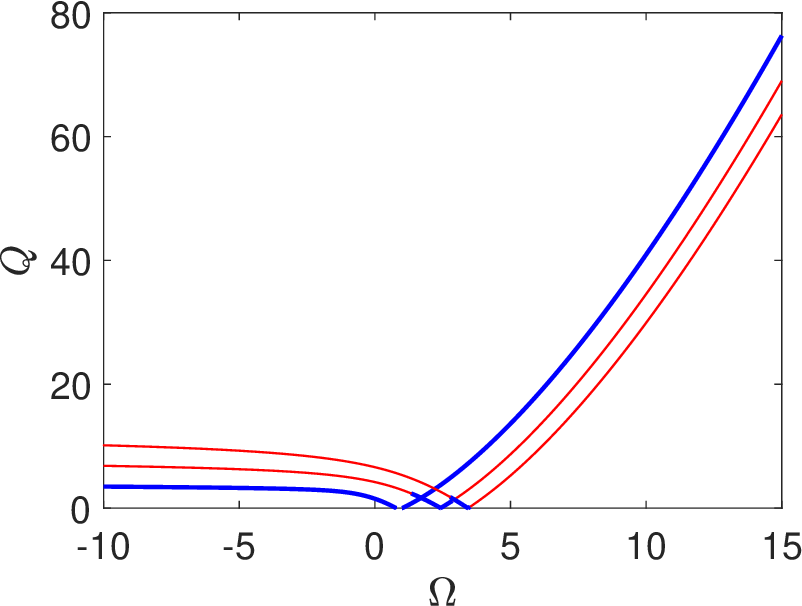}\label{subfig:bifur_alpha=1_1}}
		\caption{Bifurcation diagrams showing the $L^2$-norm $Q$ as a function of $\Omega$ for different values of the fractional parameter $\alpha$. Stable branches are shown in blue, unstable ones in red. The focusing regime corresponds to the branches extending to negative $\Omega$, while the defocusing regime corresponds to the upward branches bifurcating on the right.}
		\label{fig:bifur_alpha}
	\end{figure*}
	\begin{figure*}[htbp]
		\centering
		\subfloat[$\Omega=-2,-10$]{\includegraphics[scale=0.2]{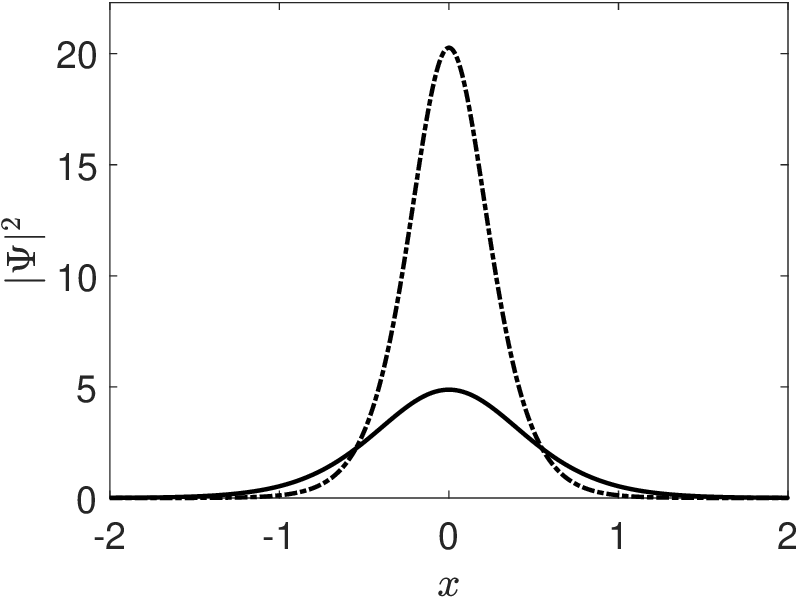}}
		\subfloat[$\Omega=-2,-10$]{\includegraphics[scale=0.2]{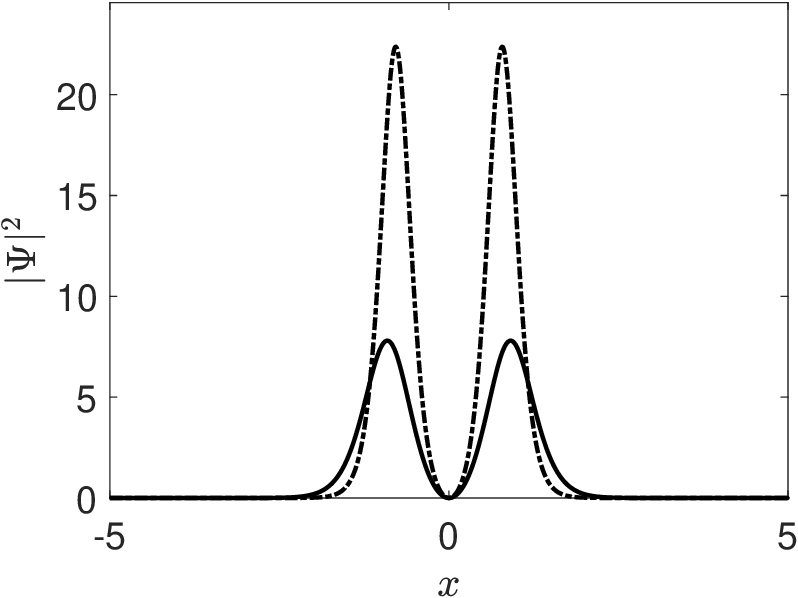}}
		\subfloat[$\Omega=-2,-10$]{\includegraphics[scale=0.2]{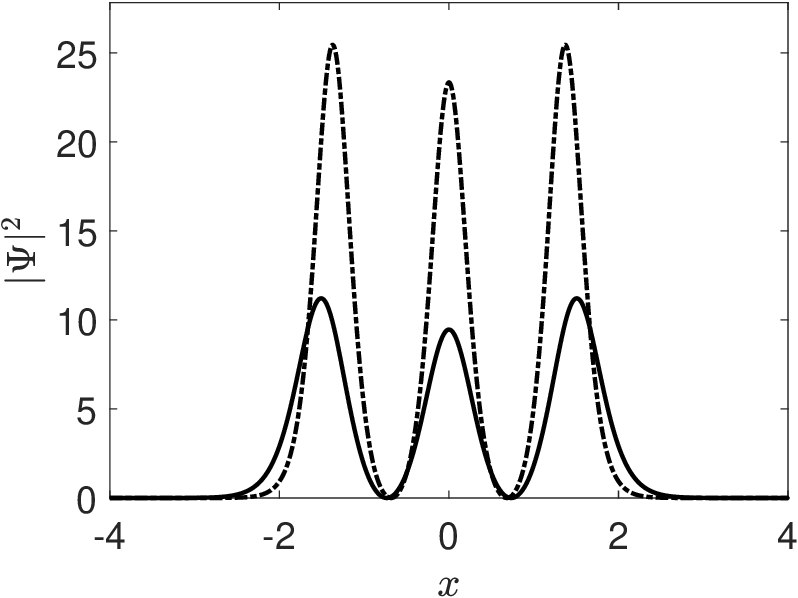}}
		\subfloat[$\Omega=7,15$]{\includegraphics[scale=0.2]{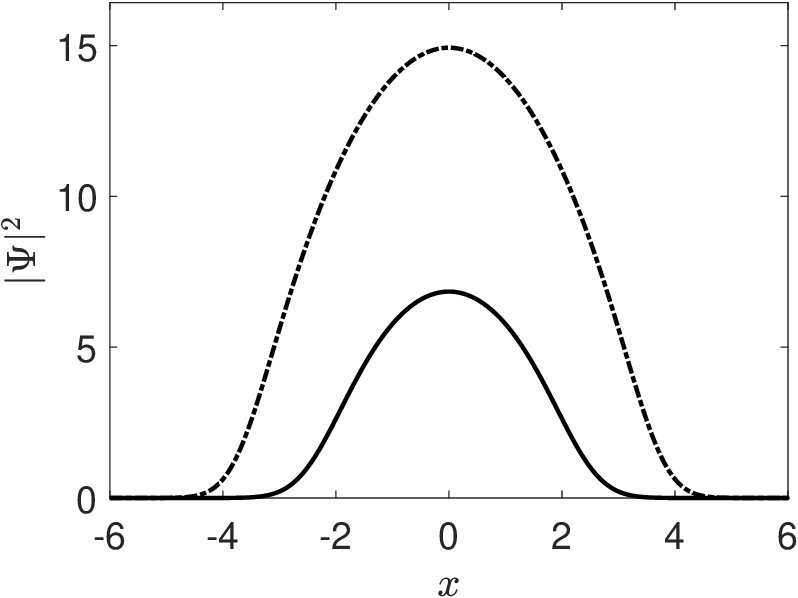}}
		\subfloat[$\Omega=7,15$]{\includegraphics[scale=0.2]{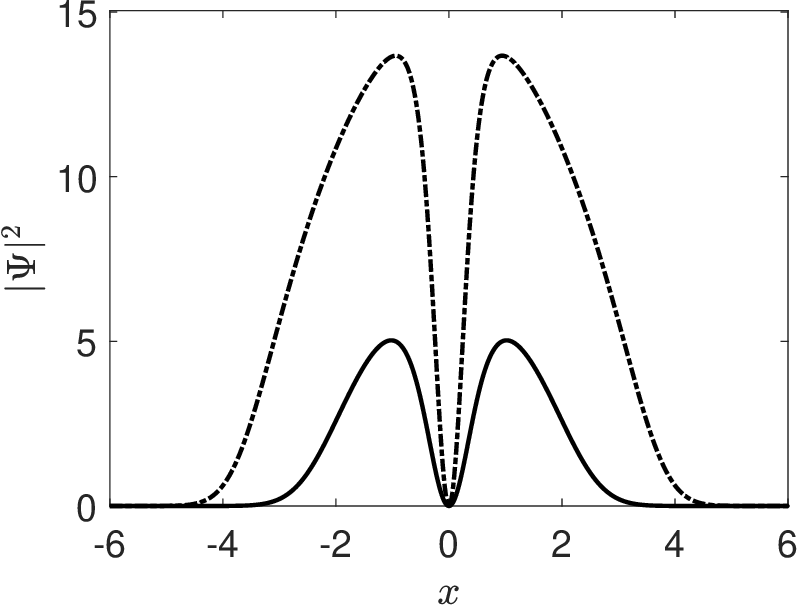}}
		\subfloat[$\Omega=7,15$]{\includegraphics[scale=0.2]{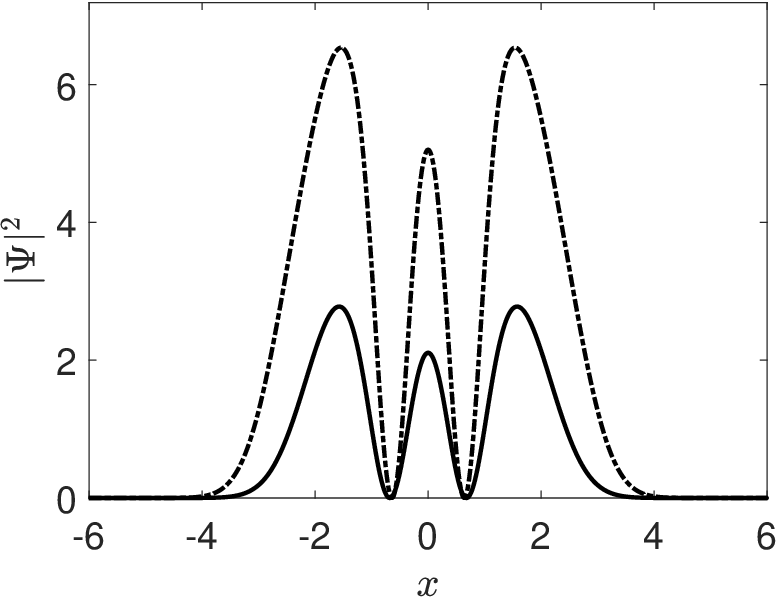}}\\
		\subfloat[$\Omega=-2,-10$]{\includegraphics[scale=0.2]{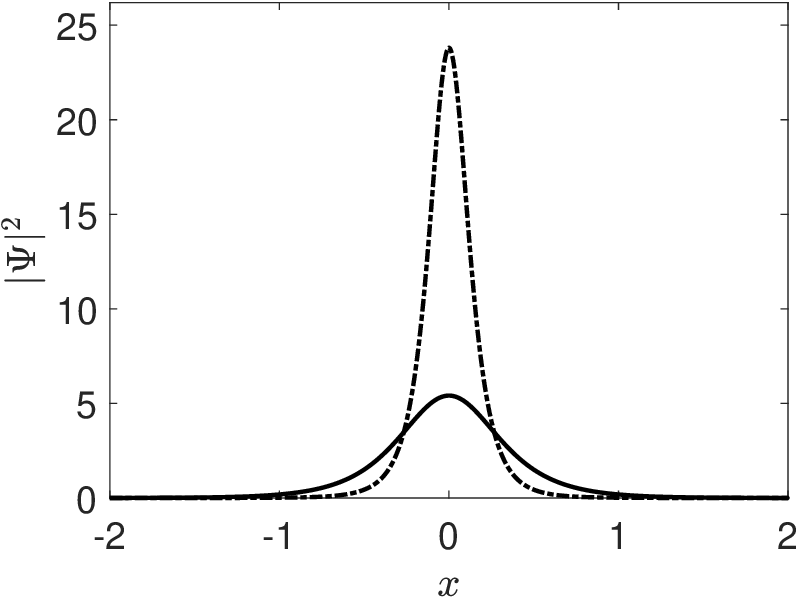}}
		\subfloat[$\Omega=-2,-10$]{\includegraphics[scale=0.2]{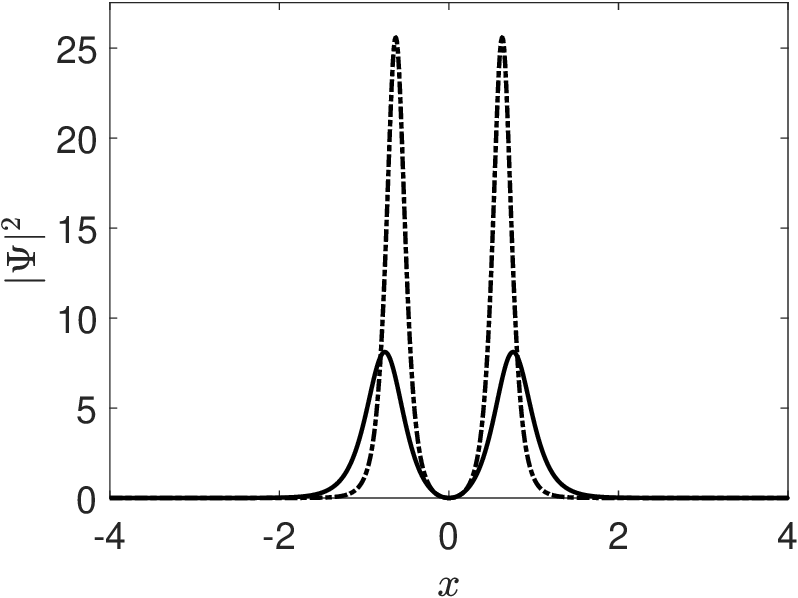}}
		\subfloat[$\Omega=-2,-10$]{\includegraphics[scale=0.2]{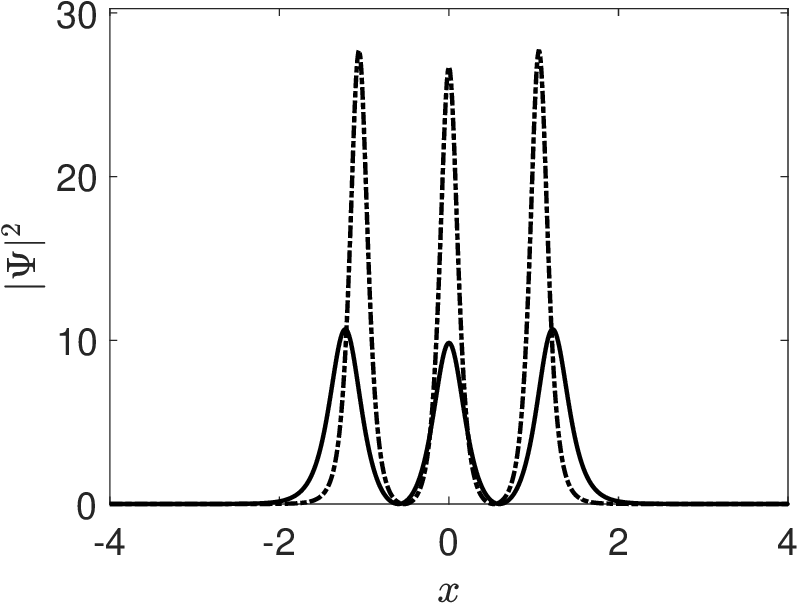}}
		\subfloat[$\Omega=7,15$]{\includegraphics[scale=0.2]{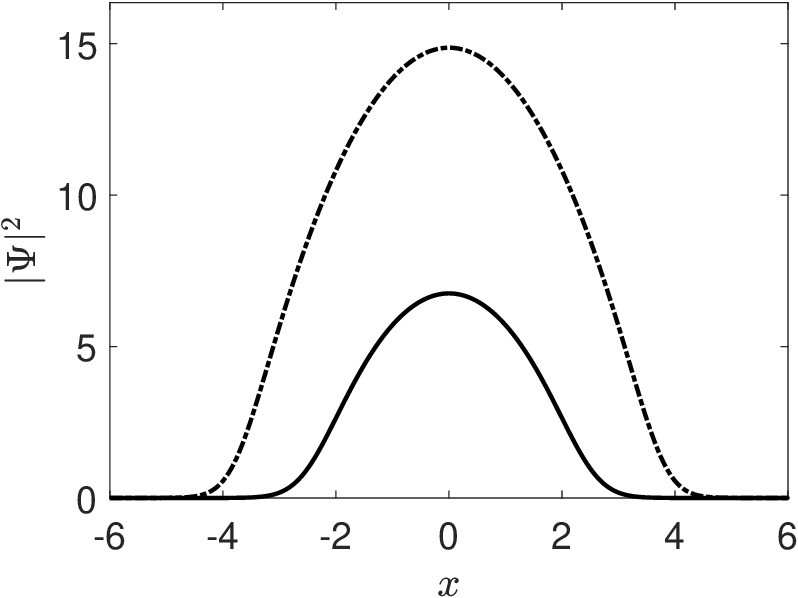}}
		\subfloat[$\Omega=7,15$]{\includegraphics[scale=0.2]{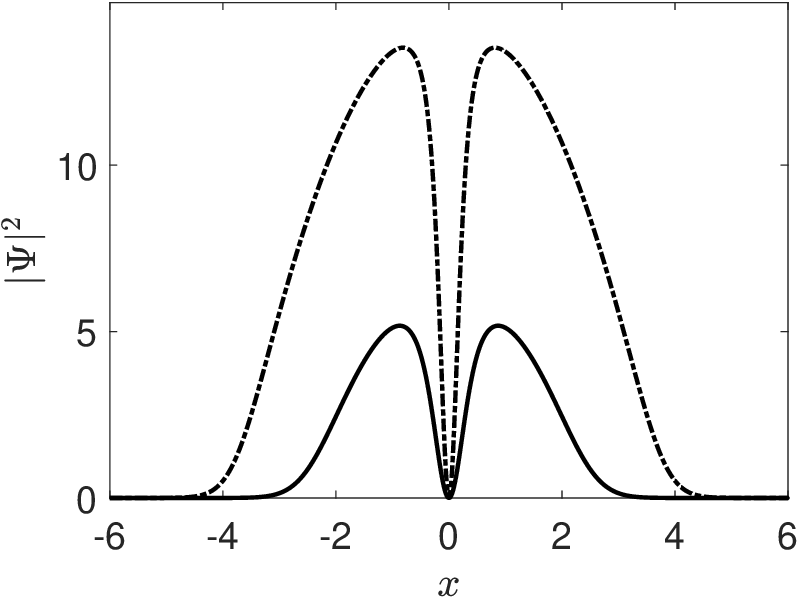}}
		\subfloat[$\Omega=7,15$]{\includegraphics[scale=0.2]{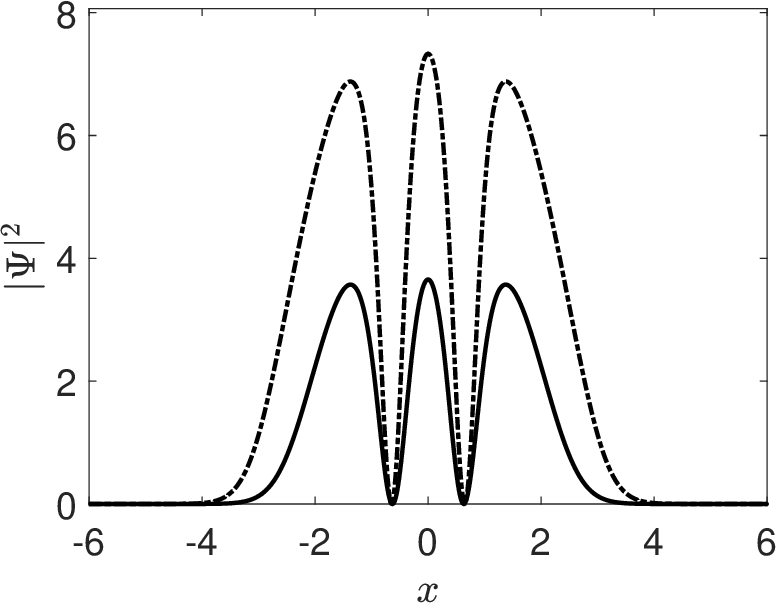}}\\
		\subfloat[$\Omega=-2,-10$]{\includegraphics[scale=0.2]{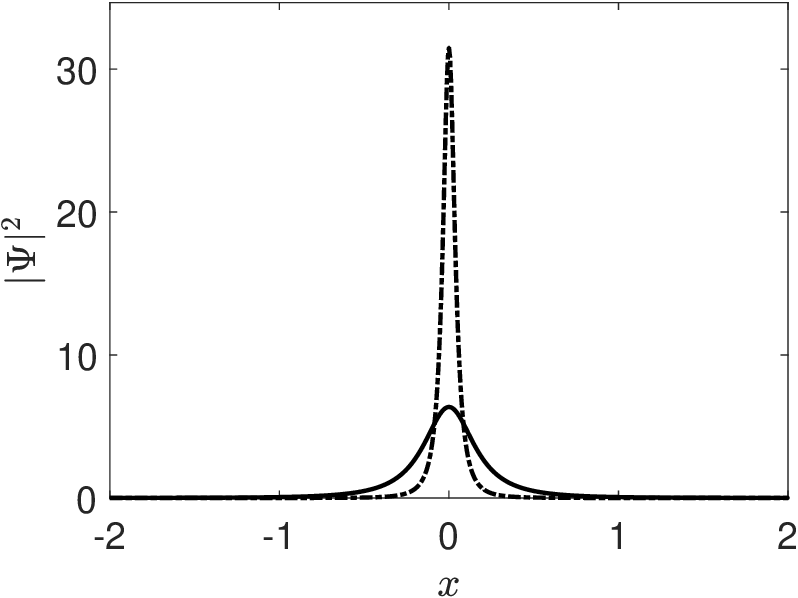}}
		\subfloat[$\Omega=-2,-10$]{\includegraphics[scale=0.2]{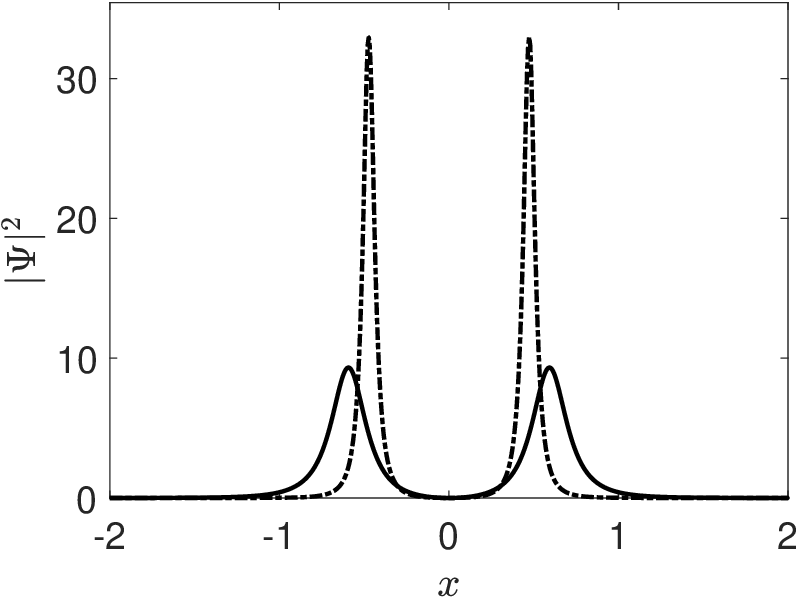}}
		\subfloat[$\Omega=-2,-10$]{\includegraphics[scale=0.2]{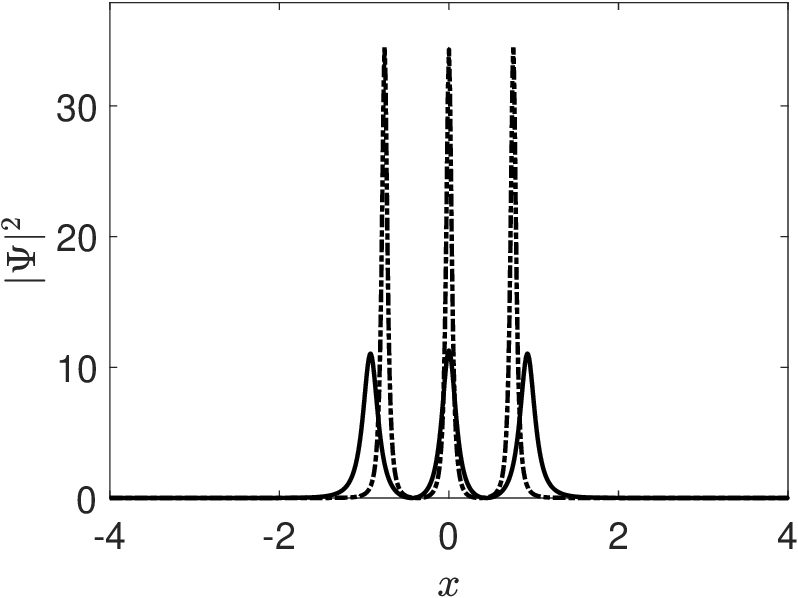}}
		\subfloat[$\Omega=7,15$]{\includegraphics[scale=0.2]{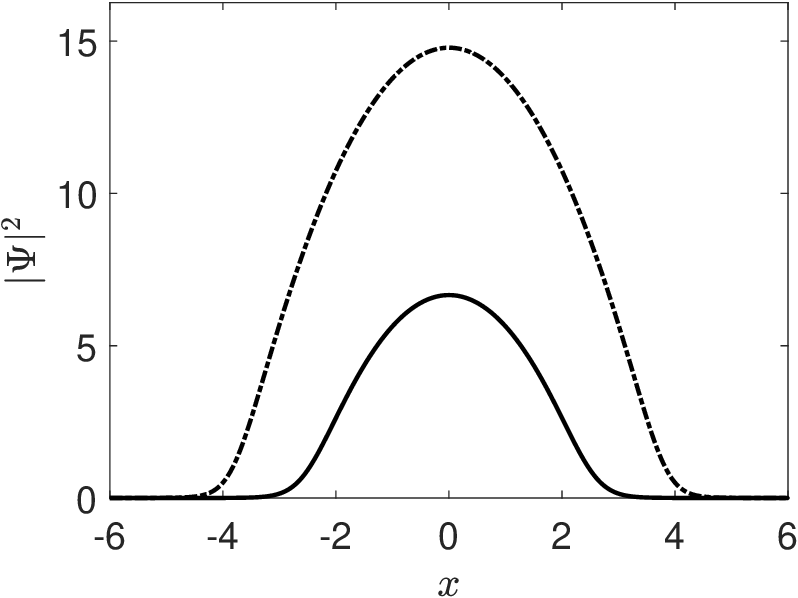}}
		\subfloat[$\Omega=7,15$]{\includegraphics[scale=0.2]{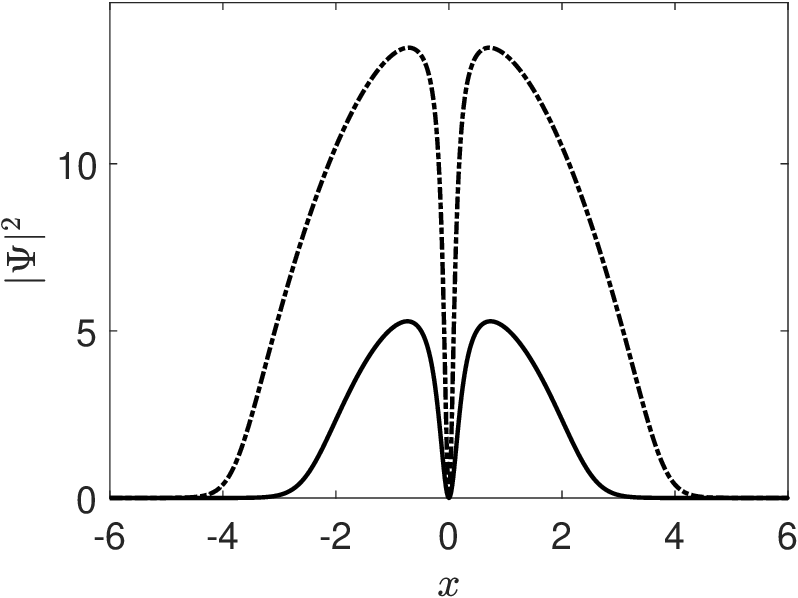}}
		\subfloat[$\Omega=7,15$]{\includegraphics[scale=0.2]{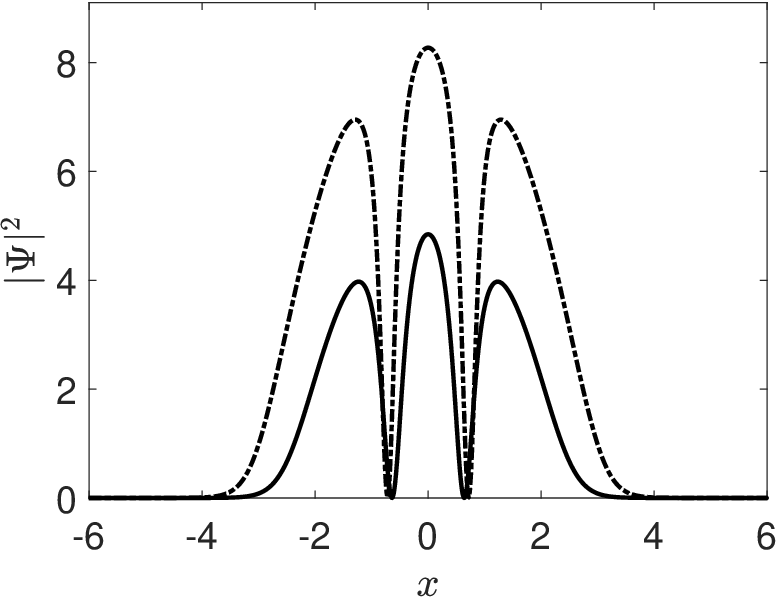}}
		\caption{Representative stationary solution profiles for $\alpha=2$ (top row), $\alpha=1.5$ (middle row), and $\alpha=1.1$ (bottom row).  
			Within each group, the first three panels correspond to focusing solutions with $\Omega=-2,-10$ for modes $n=0,1,2$, and the next three to defocusing solutions with $\Omega=7,15$ for the same modes.  
			Decreasing $\alpha$ enhances nonlocal effects: focusing states become thinner and more sensitive to $\Omega$, while defocusing states broaden and develop irregular structures compared to the classical case $\alpha=2$.}
		\label{fig:prof_alpha}
	\end{figure*}
	
	\begin{figure*}[htbp]
		\centering
		\includegraphics[scale=0.55]{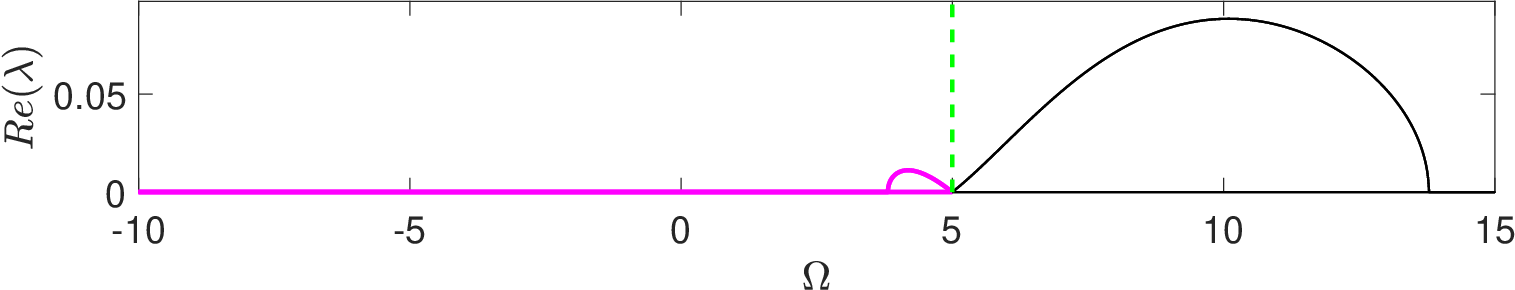}
		\caption{Real spectrum for mode $n=2$ at $\alpha=2$ in focusing (magenta, thick) and defocusing (black, thin) cases. Stability changes occur when $Re(\lambda)$ crosses zero. The green dashed line indicates the value of $\Omega$, at which the nonlinear states bifurcate.
		}
		\label{fig:spectrum_Re_alpha=2}
	\end{figure*}

	\begin{figure*}[htbp]
		\centering
		\subfloat[$n=1$]{\includegraphics[scale=0.55]{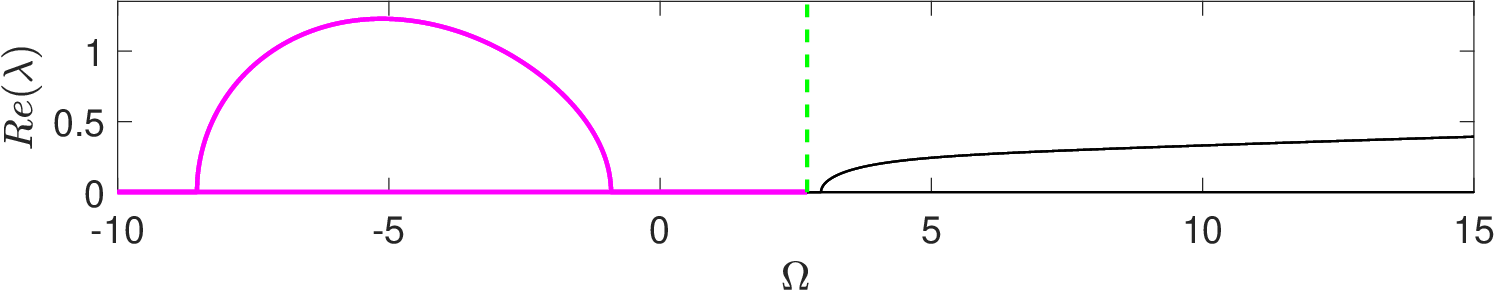}\label{subfig:spectrum_Re_alpha=1_5_mode_1}}\\
		\subfloat[$n=2$]{\includegraphics[scale=0.55]{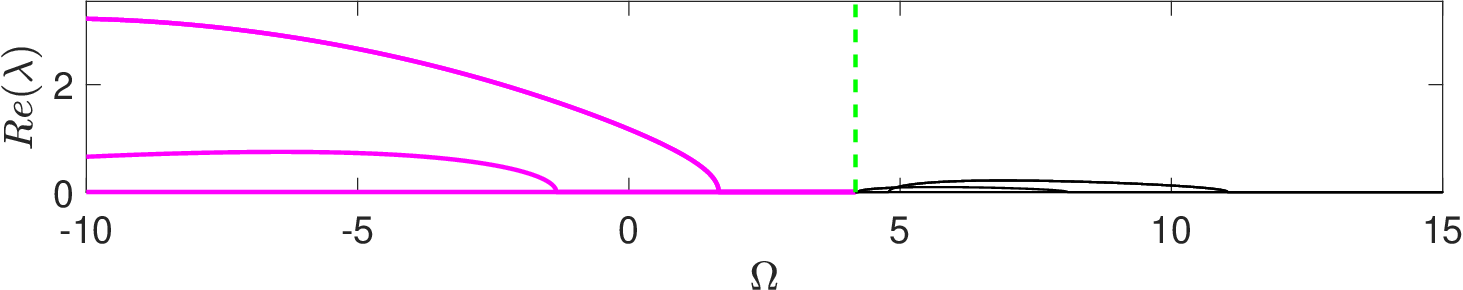}\label{subfig:spectrum_Re_alpha=1_5_mode_2}}
		\caption{Real spectrum for $\alpha=1.5$ in focusing (magenta, thick) and defocusing (black, thin) cases. Moderate nonlocality shifts bifurcation points and introduces additional unstable intervals, especially for $n=2$. 
		}
		\label{fig:spectrum_Re_alpha=1_5}
	\end{figure*}
	
	\begin{figure*}[htbp]
		\centering
		\subfloat[$n=1$]{\includegraphics[scale=0.55]{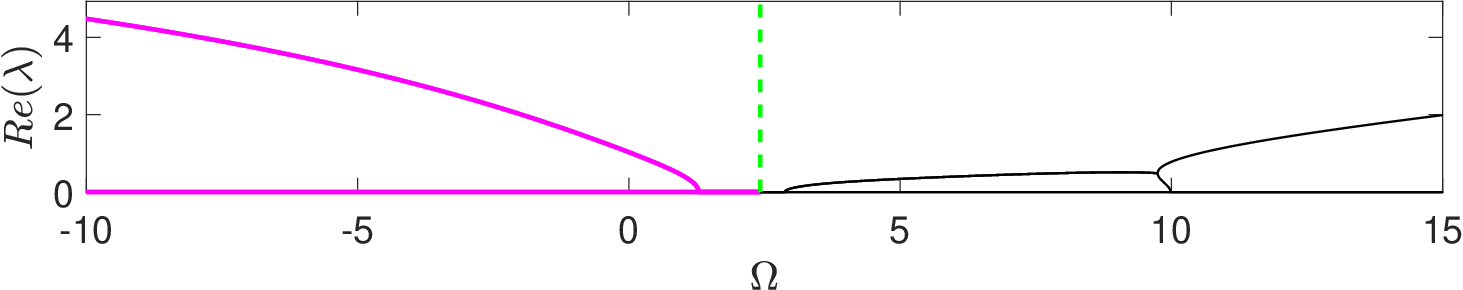}\label{subfig:spectrum_Re_alpha=1_1_mode_1}}\\
		\subfloat[$n=2$]{\includegraphics[scale=0.55]{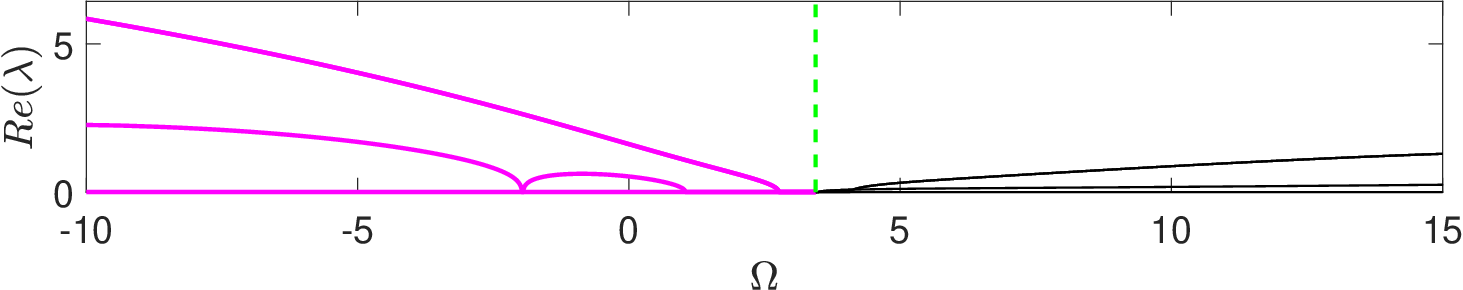}\label{subfig:spectrum_Re_alpha=1_1_mode_2}}
		\caption{Real spectrum for $\alpha=1.1$ in focusing (magenta, thick) and defocusing (black, thin) cases. Strong nonlocality fragments stability windows and enlarges unstable regions. 
		}
		\label{fig:spectrum_Re_alpha=1_1}
	\end{figure*}
	
	\begin{figure*}[htbp]
		\centering
		\subfloat[$n=0$]{\includegraphics[scale=0.55]{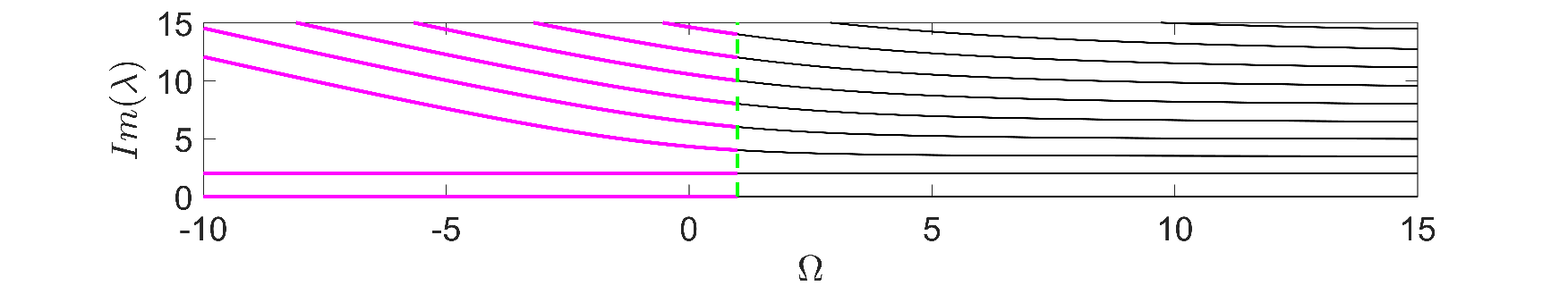}\label{subfig:spectrum_Im_alpha=2_mode_0}}\\
		\subfloat[$n=1$]{\includegraphics[scale=0.55]{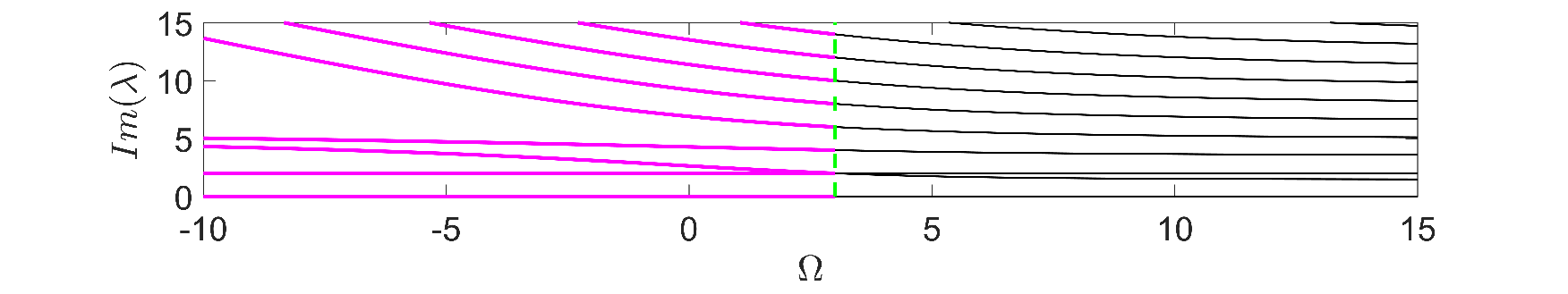}\label{subfig:spectrum_Im_alpha=2_mode_1}}\\
		\subfloat[$n=2$]{\includegraphics[scale=0.55]{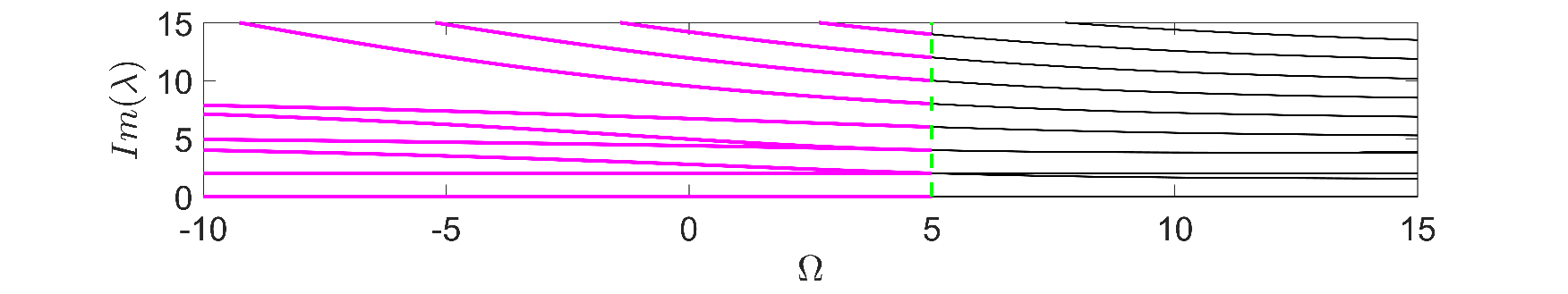}\label{subfig:spectrum_Im_alpha=2_mode_2}}
		\caption{Imaginary spectrum for $\alpha=2$ and modes $n=0,1,2$ in focusing (magenta, thick) and defocusing (black, thin) cases. The magnitude of $Im(\lambda)$ measures oscillatory content. 
		}
		\label{fig:spectrum_Im_alpha=2}
	\end{figure*}
	
	\begin{figure*}[htbp]
		\centering
		\subfloat[$n=0$]{\includegraphics[scale=0.55]{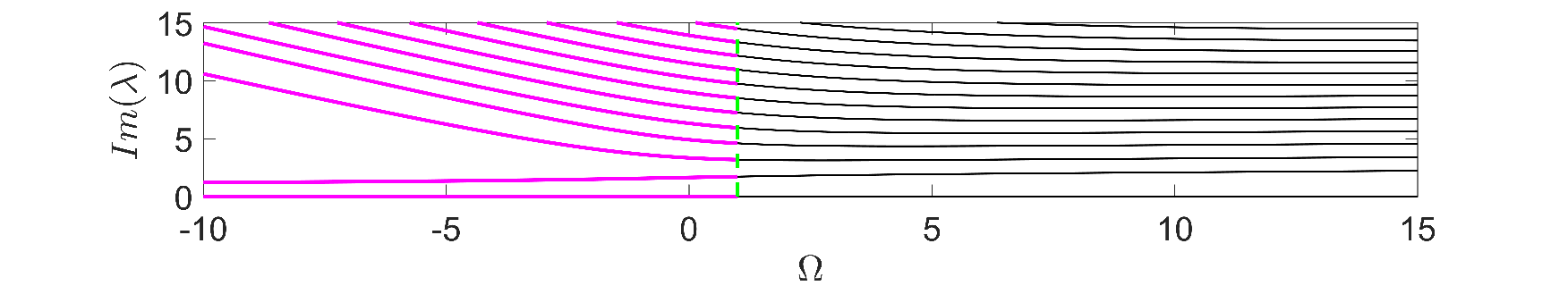}\label{subfig:spectrum_Im_alpha=1_5_mode_0}}\\
		\subfloat[$n=1$]{\includegraphics[scale=0.55]{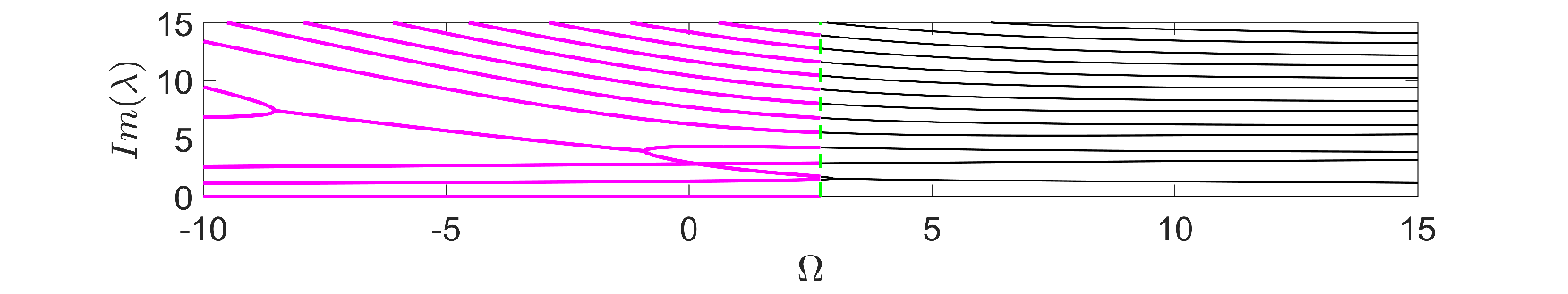}\label{subfig:spectrum_Im_alpha=1_5_mode_1}}\\
		\subfloat[$n=2$]{\includegraphics[scale=0.55]{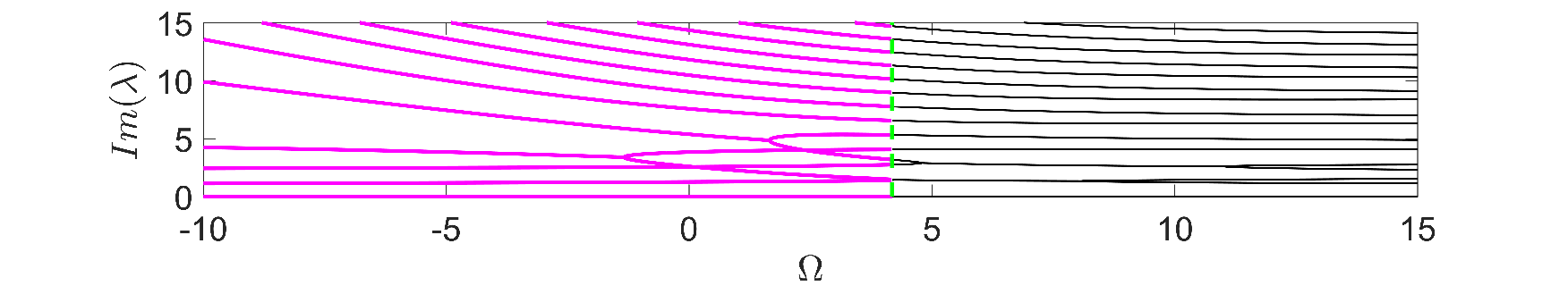}\label{subfig:spectrum_Im_alpha=1_5_mode_2}}
		\caption{Imaginary spectrum for $\alpha=1.5$. Oscillatory components strengthen with mode number and mirror the added complexity seen in the real spectrum. 
		}
		\label{fig:spectrum_Im_alpha=1_5}
	\end{figure*}
	
	\begin{figure*}[htbp]
		\centering
		\subfloat[$n=0$]{\includegraphics[scale=0.55]{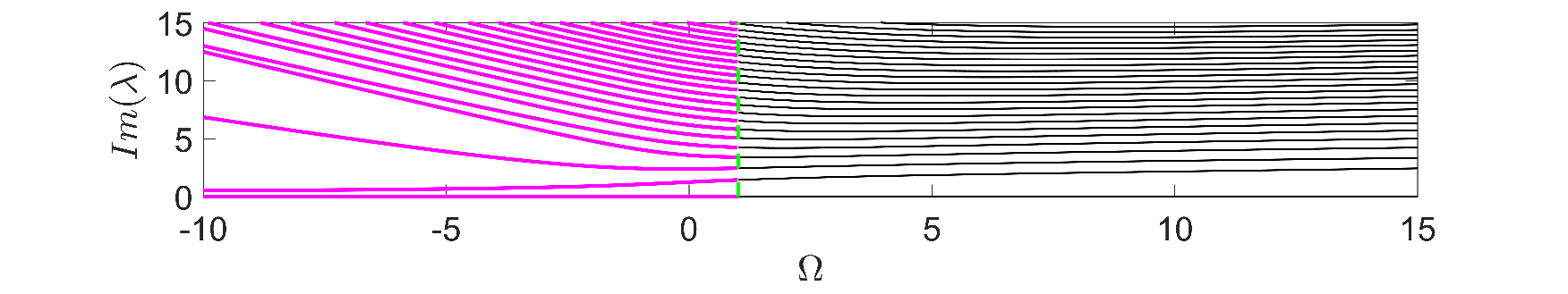}\label{subfig:spectrum_Im_alpha=1_1_mode_0}}\\
		\subfloat[$n=1$]{\includegraphics[scale=0.55]{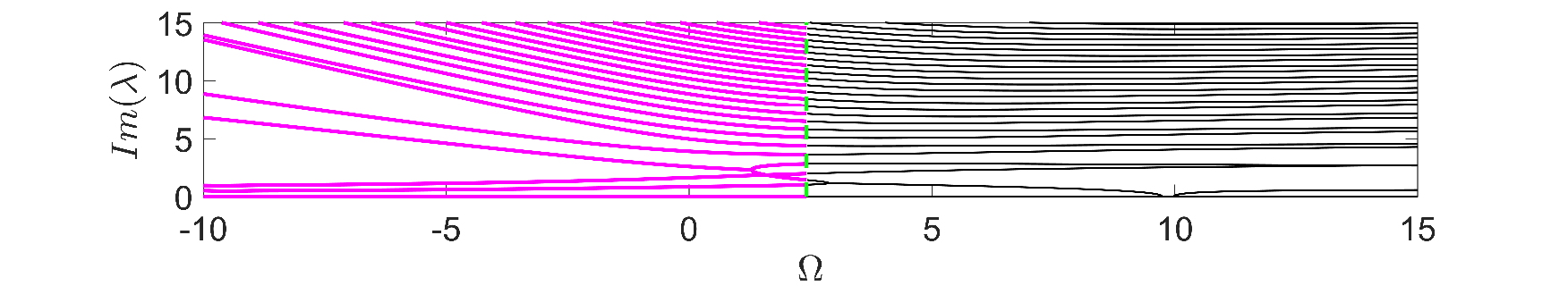}\label{subfig:spectrum_Im_alpha=1_1_mode_1}}\\
		\subfloat[$n=2$]{\includegraphics[scale=0.55]{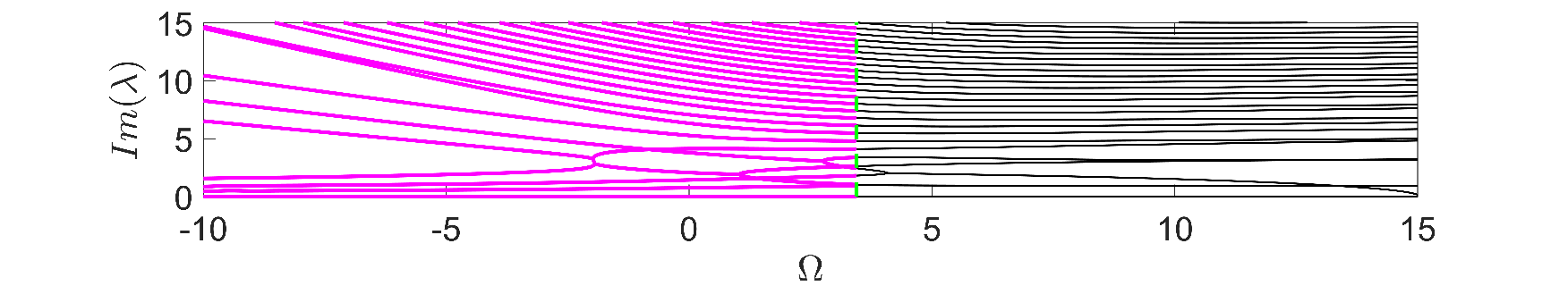}\label{subfig:spectrum_Im_alpha=1_1_mode_2}}
		\caption{Imaginary spectrum for $\alpha=1.1$. Strong nonlocality produces irregular oscillation bands across all modes, most visible for $n=2$. 
		}
		\label{fig:spectrum_Im_alpha=1_1}
	\end{figure*}

	To complement the spectral stability analysis, we examine the time evolution of \eqref{eq:fnls} using direct numerical integration. The simulations are carried out with an explicit Runge-Kutta scheme and a sufficiently small time step $dt$ to resolve nonlinear dispersive dynamics accurately. Step-size refinement tests are used to confirm that the numerical results are insensitive to further reduction of $dt$.
	
	Several diagnostics are monitored to quantify the behavior of perturbed stationary states and to connect spectral predictions with nonlinear dynamics. The first diagnostic is the \emph{mass}
	\begin{equation}
		m(t) = \int_{-\infty}^{\infty} |\psi(x,t)|^2 \, dx,
		\label{eq:mass}
	\end{equation}
	which is conserved by the continuous model. In simulations, deviations of $m(t)$ from its initial value serve as a primary measure of numerical accuracy. The \emph{momentum}
	\[
	M_x(t) = \int_{-\infty}^{\infty} x\,|\psi(x,t)|^2 \, dx
	\]
	describes the spatial distribution of mass, while 
	the \emph{center of mass}
	\begin{equation}
		X(t) = \frac
		{M_x(t)}{m(t)},
		\label{eq:centerofmass}
	\end{equation}
	tracks the average position of the wave packet. Together, these quantities indicate whether the wave packet drifts or remains localized.
	To capture structural changes in the profile, we compute the \emph{deformation measure}
	\begin{equation}
		M(t) = \left( \int_{-\infty}^{\infty} \Big( |\psi_g(x-X(t))| - |\psi(x,t)| \Big)^2 dx \right)^{1/2},
		\label{eq:deformation}
	\end{equation}
	where $\psi_g$ is a reference ground state centered at $X(t)$. This diagnostic quantifies deviations such as breathing oscillations or fragmentation relative to a localized reference state \cite{kirkpatrick2016fractional}.
	Finally, the \emph{variance}
	\begin{equation}
		S(t) = \int_{-\infty}^{\infty} (x - X(t))^2 |\psi(x,t)|^2 dx,
		\label{eq:variance}
	\end{equation}
	measures the spatial spread of the wave packet around its center of mass. Increasing $S(t)$ signals dispersive spreading, while bounded oscillations correspond to stable localization.
	
	By combining these diagnostics, i.e., mass conservation, center of mass, deformation, and variance, we obtain a clear picture of how solutions evolve under fractional dispersion. Stable stationary states remain close to their initial profile with only small oscillations, whereas unstable states exhibit growth consistent with the eigenvalue analysis, leading to spreading, deformation, or decoherence depending on $\alpha$ and $\Omega$. This diagnostic framework follows the approach of Kirkpatrick and Zhang~\cite{kirkpatrick2016fractional} and provides a robust link between spectral stability and nonlinear dynamics in fractional Schr\"odinger systems.
	
	\section{Stationary States}\label{sec:stat}	
	\subsection{Bifurcation diagrams and solution profiles}
	We present bifurcation diagrams and representative solution profiles for the ground state and the first two excited states ($n=0,1,2$) in both focusing and defocusing regimes.
	The bifurcation diagrams in Fig.~\ref{fig:bifur_alpha} show the $L^2$-norm $Q$ defined in \eqref{eq:Q_norm} as a function of the frequency parameter $\Omega$. Stable branches are indicated in blue, while unstable ones are shown in red. The left-hand branches that extend to negative $\Omega$ correspond to the focusing case ($\sigma=+1$), while the right-hand branches that bifurcate upward correspond to the defocusing case ($\sigma=-1$). These diagrams highlight how the stability structure changes with the fractional order $\alpha$. {We emphasize that focusing ($\sigma=+1$) and defocusing ($\sigma=-1$) cases correspond to different equations and are treated separately. For a fixed value of $\sigma$, stationary branches of Eq.~\eqref{eq:stat} bifurcate from the linear eigenvalues only on one side of $\Omega$. On the opposite side, the stationary problem does not support localized nontrivial solutions, and numerical continuation initialized from the linear modes converges only to the trivial state $\tilde{\Phi}\equiv 0$.
	}
	
	For $\alpha=2$ (classical Laplacian), the bifurcation curves are regular and align with well-known results for local nonlinear Schr\"odinger systems: stability transitions occur at predictable values of $\Omega$. When $\alpha=1.5$, moderate nonlocality shifts bifurcation points and produces additional stability transitions, with branches alternating between stable and unstable segments. At $\alpha=1.1$, strong nonlocality dominates, giving rise to closely spaced bifurcations and additional branches, resulting in a more intricate stability diagram.

	The solution profiles in Fig.~\ref{fig:prof_alpha} provide a detailed picture of how fractional dispersion modifies the shape of stationary states. For $\alpha=2$ (the classical Laplacian), the focusing states at $\Omega=-2$ and $\Omega=-10$ (panels a-c) are broad but symmetric, with smooth decay and well-defined nodal structure for modes $n=0,1,2$. In contrast, the defocusing states at $\Omega=7$ and $\Omega=15$ (panels d-f) expand outward: the ground state becomes wider, and higher modes flatten, reflecting the spreading effect typical of local defocusing nonlinear Schr\"odinger dynamics.  
	
	When $\alpha$ decreases to $1.5$, moderate nonlocality alters these trends. The focusing profiles (g-i) become thinner and more sharply localized compared to $\alpha=2$, while their amplitudes increase slightly. This sharpening indicates that nonlocal interactions strengthen localization, but also make the solutions more sensitive to changes in $\Omega$, as seen from the noticeable variations between $\Omega=-2$ and $\Omega=-10$. On the other hand, the defocusing profiles (j-l) broaden more strongly than in the classical case, with flatter peaks and less regular shapes, showing that nonlocal effects enhance the dispersive character of defocusing solutions.  
	
	For $\alpha=1.1$, strong nonlocality dominates and produces striking structural changes. The focusing states (m-o) are still localized, but now they are very thin and sharply peaked, with amplitudes that react strongly to small changes in $\Omega$. The higher modes ($n=1,2$) show narrower lobes and reduced symmetry, highlighting the fragility of excited states under strong nonlocal interactions. In the defocusing regime (p-r), the solutions broaden dramatically, with profiles that lose their regular parabolic shape and develop irregular or flattened structures. These features signal the destabilizing influence of long-range interactions, which spread mass more unevenly across space.  
	
	Overall, decreasing $\alpha$ enhances nonlocal effects and drives a clear contrast between focusing and defocusing regimes. In focusing dynamics, smaller $\alpha$ values produce thinner and more sensitive localized states, while in defocusing dynamics, they lead to broader and less regular profiles. Compared to the classical case $\alpha=2$, the profiles at $\alpha=1.5$ and $\alpha=1.1$ demonstrate how fractional dispersion reshapes both the width and stability of stationary states, with the strongest effects observed at low $\alpha$.

	\subsection{Spectrum analysis}
	
	Figures~\ref{fig:spectrum_Re_alpha=2}-\ref{fig:spectrum_Im_alpha=1_1} display how the eigenvalue spectrum depends on the fractional order $\alpha$ for modes $n=0,1,2$ in both focusing ($\sigma=+1$) and defocusing ($\sigma=-1$) regimes. The real part $Re(\lambda)$ governs stability: $Re(\lambda)>0$ indicates growth of perturbations, while $Re(\lambda)\leq0$ corresponds to spectral stability. The imaginary part $Im(\lambda)$ reflects oscillatory dynamics of the linearized modes. The vertical green dashed line in each figure marks the bifurcation boundary in $\Omega$ beyond which nonlinear modes emerge. {For each fixed value of $\sigma$, the Lyapunov bands $\lambda(\Omega)$ vary continuously along the corresponding stationary branch and terminate at the linear bifurcation point, consistent with the bifurcation structure shown in Fig.~\ref{fig:bifur_alpha}.
	}

	Figure~\ref{fig:spectrum_Re_alpha=2} shows the case $\alpha=2$ (classical Laplacian). For mode $n=2$, the defocusing branch (black, thin) remains unstable over a wide range of $\Omega$, while the focusing branch (magenta, thick) exhibits a narrow instability window that terminates at the dashed line. This matches the bifurcation structure seen in the $Q(\Omega)$ curves.  
	At $\alpha=1.5$ (Fig.~\ref{fig:spectrum_Re_alpha=1_5}), moderate nonlocality shifts the bifurcation point leftward, reducing the stability interval for focusing states. For $n=1$ and $n=2$, extra unstable regions appear, especially in the focusing case, where $Re(\lambda)$ rises above zero in multiple intervals.  
	At $\alpha=1.1$ (Fig.~\ref{fig:spectrum_Re_alpha=1_1}), strong nonlocality dominates: stability windows are fragmented, and both $n=1$ and $n=2$ rapidly lose stability. The defocusing case retains wider stability domains, but even here the unstable intervals broaden as $\Omega$ increases. In summary, decreasing $\alpha$ compresses stability windows, shifts bifurcation points toward lower $\Omega$, and destabilizes higher modes earlier. The focusing regime is always more fragile than the defocusing one.

	Figures~\ref{fig:spectrum_Im_alpha=2}-\ref{fig:spectrum_Im_alpha=1_1} illustrate the oscillatory content $Im(\lambda)$. For $\alpha=2$ (Fig.~\ref{fig:spectrum_Im_alpha=2}), oscillations are regular and increase smoothly with mode number: $n=0$ shows nearly flat lines, while $n=2$ exhibits a richer band structure. Importantly, the focusing (magenta) and defocusing (black) cases share similar imaginary patterns, even though their growth rates differ in the real part.  
	At $\alpha=1.5$ (Fig.~\ref{fig:spectrum_Im_alpha=1_5}), oscillatory components strengthen with mode number, and the spectrum becomes more intricate, mirroring the fragmented stability structure seen in the real part. For $n=2$, oscillations intensify near the dashed line, signaling mode interactions close to bifurcation.  
	At $\alpha=1.1$ (Fig.~\ref{fig:spectrum_Im_alpha=1_1}), strong nonlocality yields irregular oscillation bands across all modes. The higher mode $n=2$ shows the most complex structures, with oscillation branches bending and splitting near bifurcation boundaries. This highlights how fractional dispersion alters not only growth rates but also the internal oscillatory modes.

	\begin{figure*}[htbp]
		\centering
		\subfloat[$n=1$]{\includegraphics[scale=0.45]{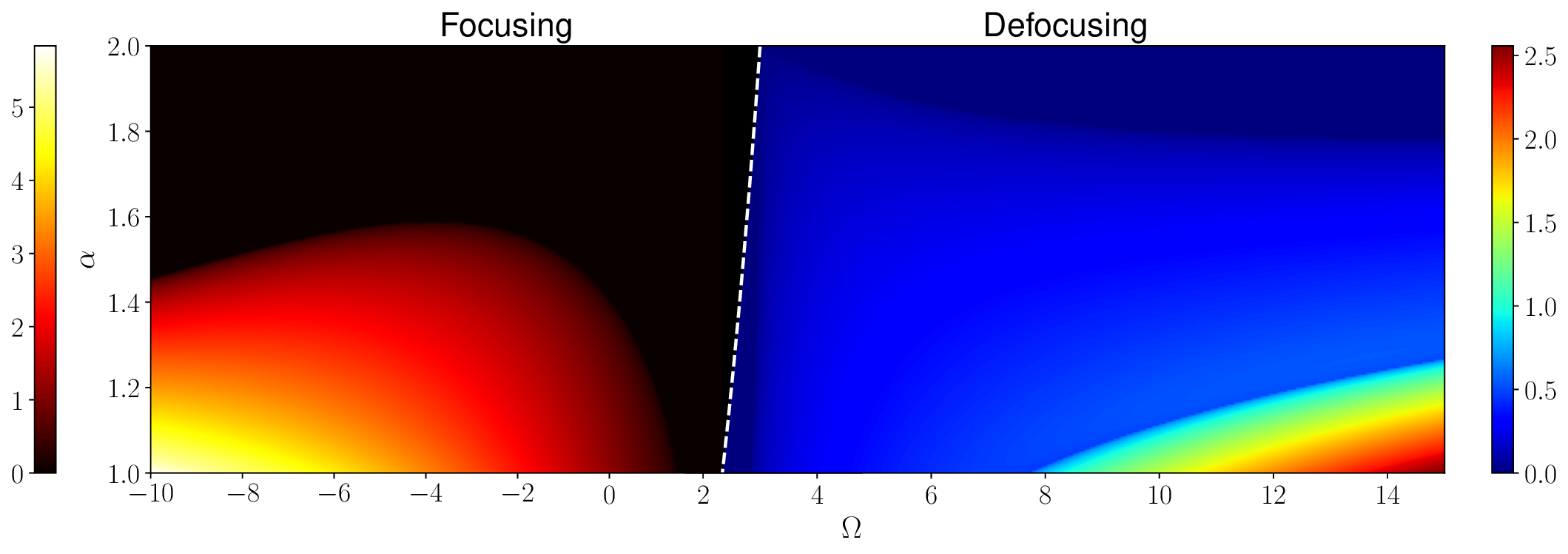}\label{subfig:alpha_vs_omega_n_1}}\\
		\subfloat[$n=2$]{\includegraphics[scale=0.45]{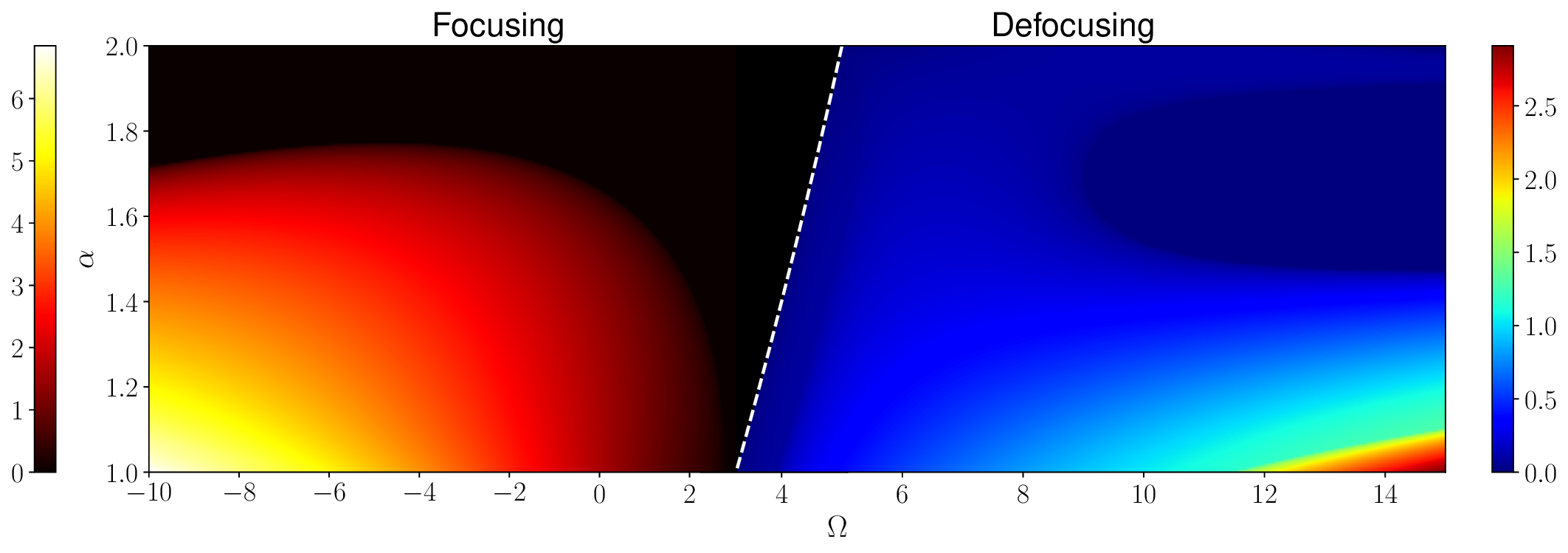}\label{subfig:alpha_vs_omega_n_2}}
		\caption{Stability maps in the $(\alpha,\Omega)$ plane for modes $n=1$ (a) and $n=2$ (b). Left: focusing; right: defocusing. The {white} dashed curve marks the bifurcation boundary obtained from the eigenvalue problem. The ground state $n=0$ remains stable for all parameters.}
		\label{fig:alpha_vs_omega}
	\end{figure*}

	Figure~\ref{fig:alpha_vs_omega} summarizes the stability regions in the $(\alpha,\Omega)$ plane. The ground state ($n=0$) is always stable, consistent with its role as the lowest-energy mode. For $n=1$ and $n=2$, stability domains shrink as $\alpha$ decreases: focusing states lose stability earlier (smaller $\Omega$) than defocusing states. The dashed lines in Fig.~\ref{fig:alpha_vs_omega} mark the exact stability boundaries computed from the linearized eigenvalue problem, confirming the bifurcation points observed in the spectra.

	In conclusion, lowering $\alpha$ destabilizes higher modes, narrows stability windows, and enhances oscillatory complexity. The focusing regime exhibits narrow and fragile stability domains, while the defocusing regime retains broader but eventually fragmented stable regions. The imaginary spectra reveal that both focusing and defocusing share similar oscillatory structures, even when their stability differs, pointing to a common dynamical backbone shaped by fractional dispersion. Compared to the classical Schr\"odinger case ($\alpha=2$), fractional models ($\alpha<2$) thus display richer but more fragile spectral behavior.
	
	\section{Time Dynamics}\label{sec:dynamics}
	
	To illustrate the nonlinear behavior predicted by the spectral analysis, we now present representative simulations of the time evolution under fractional dispersion (\(\alpha\neq 2\)). The diagnostics introduced in Section~\ref{sec:methods}--namely the mass $m(t)$, center of mass $X(t)$, variance $S(t)$, and deformation measure $M(t)$--are used to quantify stability, localization, and structural changes. Rather than repeating the definitions, we emphasize here their physical interpretation: $X(t)$ monitors translational drift, $S(t)$ measures the spread of the wave packet, and $M(t)$ detects profile deformations such as breathing or fragmentation. These quantities provide the link between spectral predictions and fully nonlinear dynamics.
	
	\subsection{Unstable evolutions}
	\begin{figure*}[tbhp!]
		\centering
		\subfloat[Spatiotemporal magnitude \(|\psi(x,t)|\)]{\includegraphics[scale=0.3]{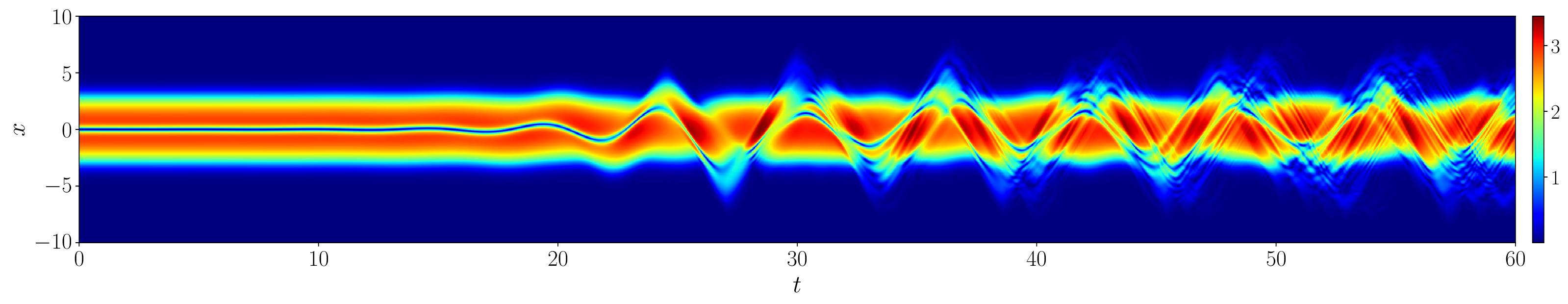}\label{subfig:defocusing_alpha=1_5_mode=1_omega=10_image_data}}\\
		\subfloat[Phase \(\arg \psi(x,t)\)]{\includegraphics[scale=0.3]{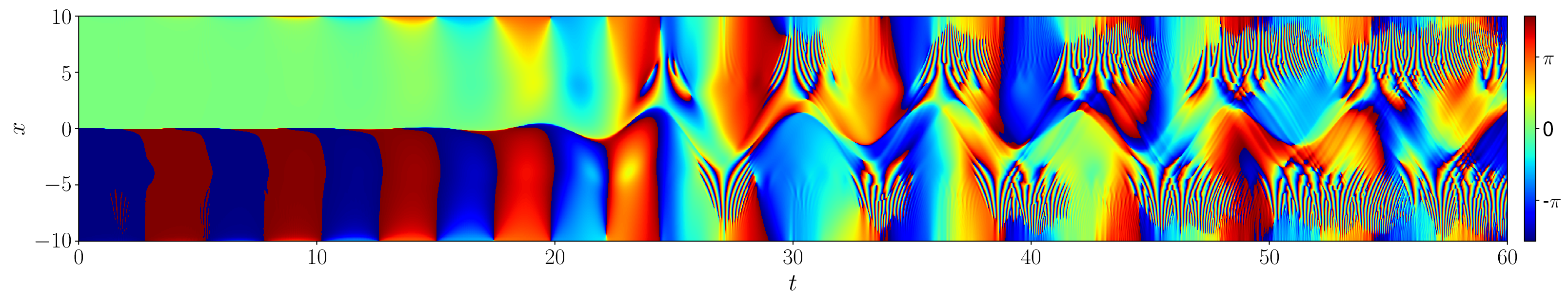}\label{subfig:defocusing_alpha=1_5_mode=1_omega=10_angle}}\\
		\subfloat[3D profile view]{\includegraphics[scale=0.55]{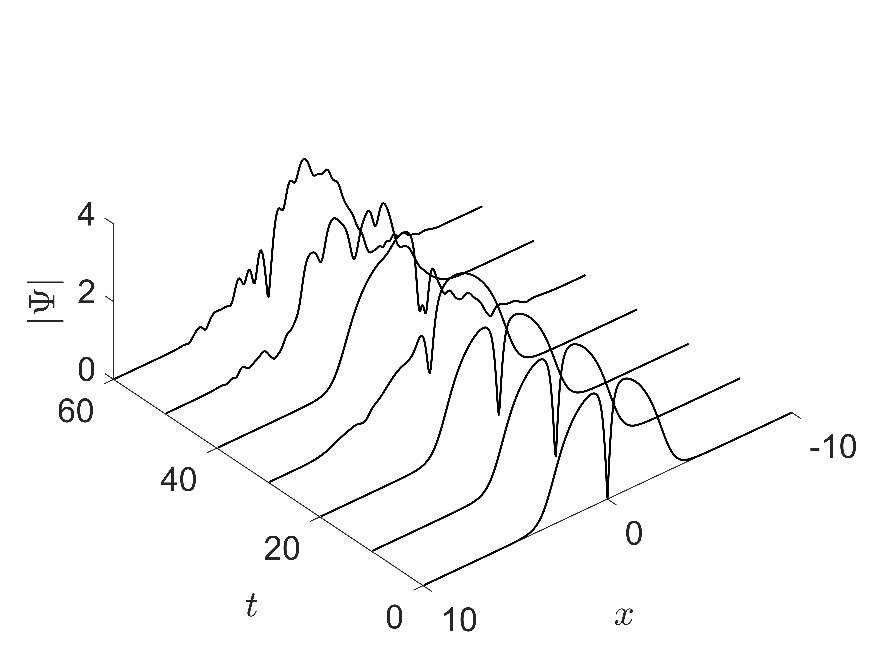}\label{subfig:defocusing_alpha=1_5_mode=1_omega=10_3D}}\quad
		\subfloat[Diagnostics \(X(t),S(t),M(t)\)]{\includegraphics[scale=0.55]{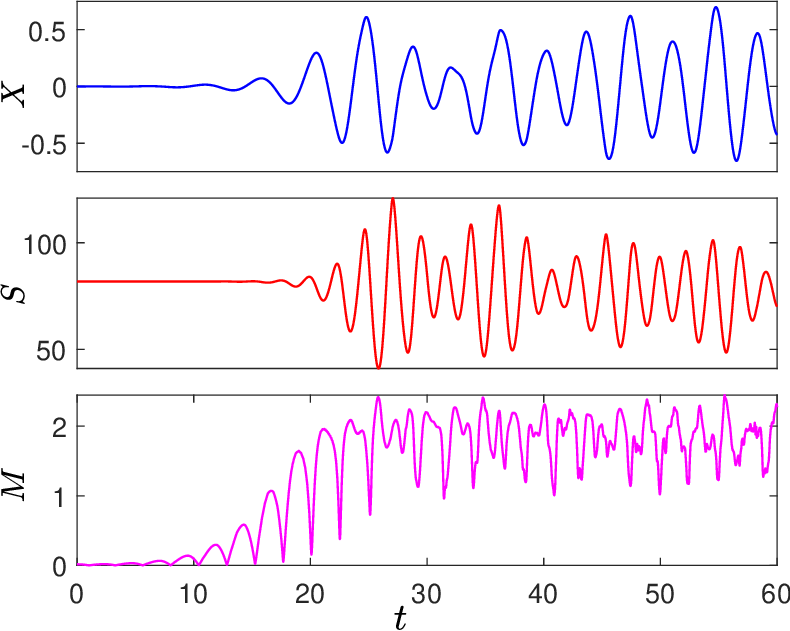}\label{subfig:defocusing_alpha=1_5_mode=1_omega=10_XSM}}
		\caption{Defocusing case, \(\alpha=1.5\), \(n=1\), \(\Omega=10\). 
		The mode is stable in the integer-order case $\alpha=2$. Shown is the typical dynamical evolution due to the instability.}
		\label{fig:defocusing_alpha=1_5_mode=1_omega=10}
	\end{figure*}

	Figure~\ref{fig:defocusing_alpha=1_5_mode=1_omega=10} presents a representative sample of the time dynamics in the defocusing regime with \(\alpha=1.5\), \(n=1\), and \(\Omega=10\). 
	The dynamics illustrate that fractional dispersion destabilizes a mode that is stable in the classical local case $\alpha=2$. The intrinsically nonlocal character of the fractional Laplacian modifies the dispersive balance underlying the mode. In the spatiotemporal evolution of the amplitude $|\psi(x,t)|$ [Fig.~\ref{subfig:defocusing_alpha=1_5_mode=1_omega=10_image_data}], the central dip develops small-amplitude oscillations while the wave packet initially remains spatially localized. At later times, however, visible deformation of the profile emerges. The phase evolution [Fig.~\ref{subfig:defocusing_alpha=1_5_mode=1_omega=10_angle}] is initially smooth and spatially coherent, but progressively exhibits irregular spatiotemporal modulations. The 3D visualization [Fig.~\ref{subfig:defocusing_alpha=1_5_mode=1_omega=10_3D}] shows both the onset of dip oscillations and the subsequent deformation of the wave packet.
	
	The quantitative diagnostics [Fig.~\ref{subfig:defocusing_alpha=1_5_mode=1_omega=10_XSM}] further corroborate this behavior. The center of mass $X(t)$ remains constant at early times, but subsequently begins to oscillate, in direct correspondence with the oscillatory motion of the dip. The variance $S(t)$, following an initial plateau, also develops oscillations, indicating breathing dynamics characterized by alternating phases of localization and delocalization. In addition, the deformation measure $M(t)$ rises sharply, signaling a progressive loss of structural invariance of the profile.
	
	This example should be understood as one sample of the possible dynamics in the unstable regime. 
	Other parameter choices may lead to stronger instabilities or complete decoherence, as will be illustrated in subsequent examples.

	\subsection{Coherence}
	
	\begin{figure*}[htbp]
		\centering
		\subfloat[Spatiotemporal magnitude \(|\psi(x,t)|\)]{\includegraphics[scale=0.3]{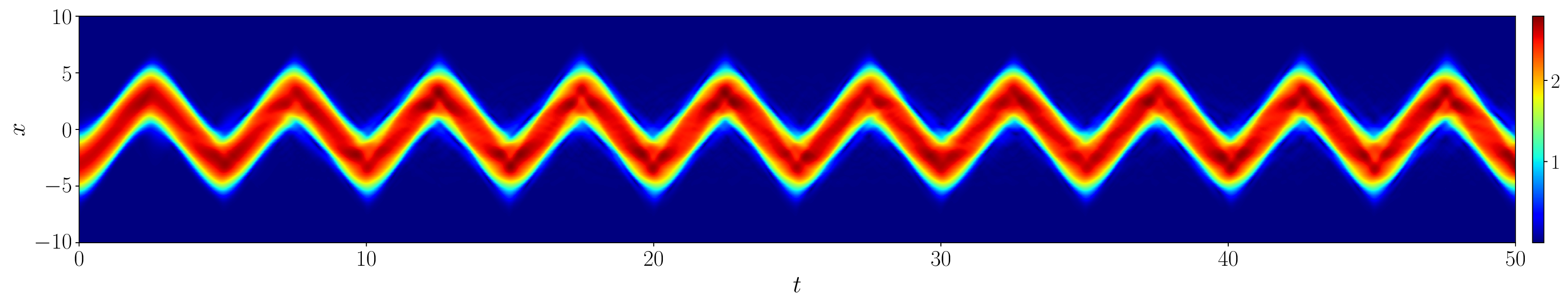}\label{subfig:defocusing_alpha=1_5_mode=0_omega=7_h=3_image_data}}\\
		\subfloat[Phase \(\arg \psi(x,t)\)]{\includegraphics[scale=0.3]{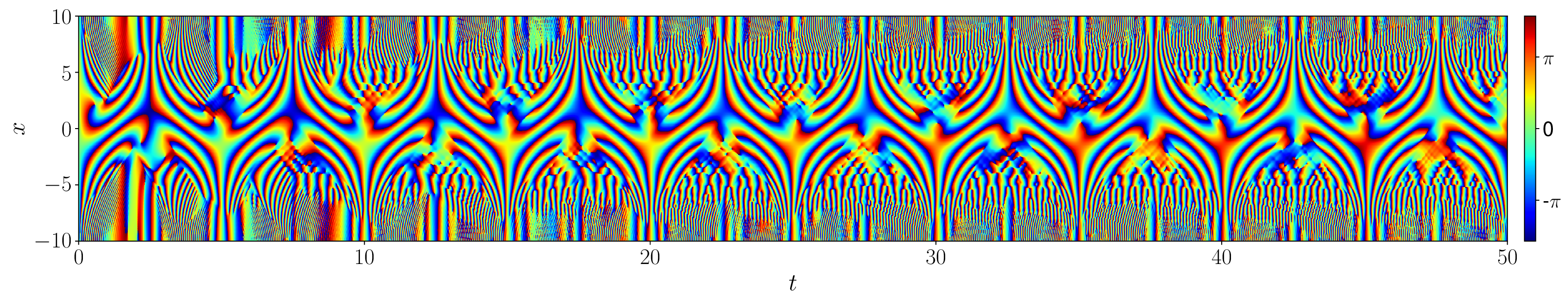}\label{subfig:defocusing_alpha=1_5_mode=0_omega=7_h=3_angle}}\\
		\subfloat[3D profile view]{\includegraphics[scale=0.55]{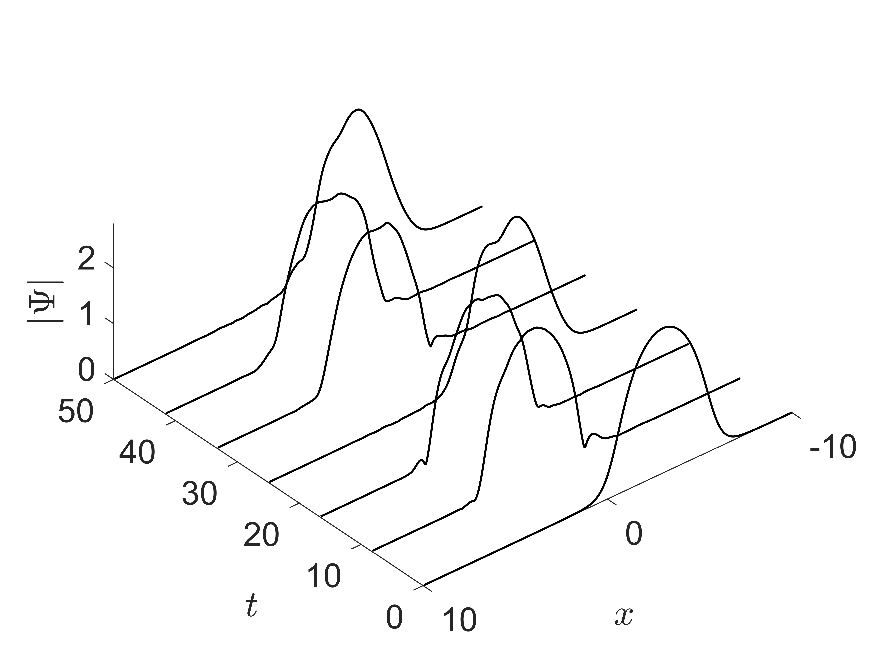}\label{subfig:defocusing_alpha=1_5_mode=0_omega=7_h=3_3D}}\quad
		\subfloat[Diagnostics \(X(t),S(t),M(t)\)]{\includegraphics[scale=0.55]{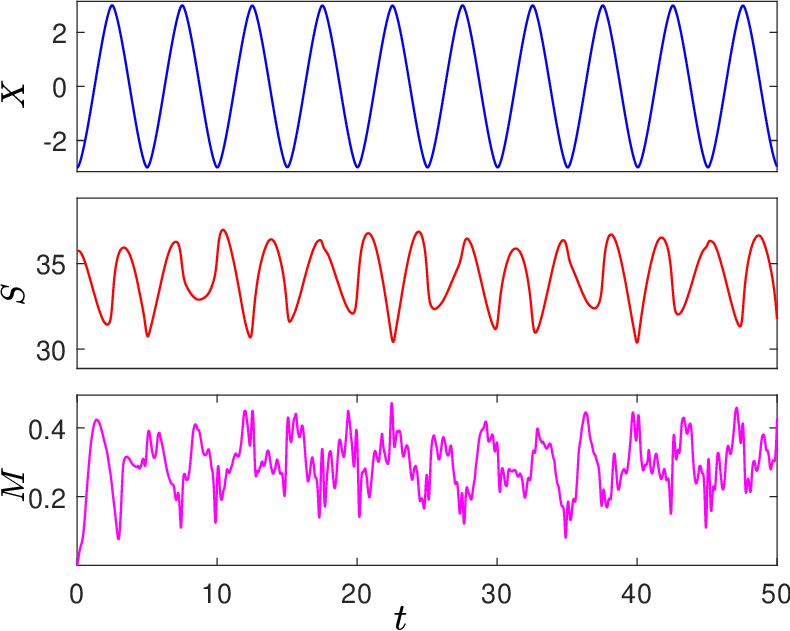}\label{subfig:defocusing_alpha=1_5_mode=0_omega=7_h=3_XSM}}
		\caption{Coherent evolution in the defocusing regime with \(\alpha=1.5\), \(n=0\), \(\Omega=7\), and shift \(h=3\). The wave packet remains localized and stable: the magnitude is regular, the phase smooth, and the diagnostics confirm coherent dynamics. This example illustrates how fractional dispersion supports robust coherence even under an initial displacement.}
		\label{fig:defocusing_alpha=1_5_mode=0_omega=7_h=3}
	\end{figure*}
	Figure~\ref{fig:defocusing_alpha=1_5_mode=0_omega=7_h=3} shows a coherent evolution in the defocusing regime for \(\alpha=1.5\), the ground state (\(n=0\)), and \(\Omega=7\) with the initial condition $\psi(x,0)=\tilde{\Phi}(x-h)$ and an initial shift \(h=3\). The spatiotemporal magnitude plot (Fig.~\ref{subfig:defocusing_alpha=1_5_mode=0_omega=7_h=3_image_data}) demonstrates that the wave packet oscillates but does not disperse, remaining well localized over long times. The phase plot (Fig.~\ref{subfig:defocusing_alpha=1_5_mode=0_omega=7_h=3_angle}) exhibits smooth, regular patterns without irregularities, consistent with internal coherence. The 3D view (Fig.~\ref{subfig:defocusing_alpha=1_5_mode=0_omega=7_h=3_3D}) confirms that the profile maintains its structure, with only mild oscillations due to the imposed spatial shift.
	
	The diagnostics (Fig.~\ref{subfig:defocusing_alpha=1_5_mode=0_omega=7_h=3_XSM}) reinforce this conclusion: \(X(t)\) oscillates periodically around the shifted center but shows no drift, indicating stable translational dynamics; \(S(t)\) varies within a narrow band, reflecting bounded breathing oscillations without spreading; and \(M(t)\) remains very small, confirming that the shape of the wave packet is preserved. 
	The overall structure remains robust due to the stabilizing effect of the defocusing nonlinearity.
	
	This case represents a typical coherent evolution under fractional dispersion: even with an imposed displacement, the wave packet remains stable, localized, and phase-coherent, highlighting the resilience of defocusing states when \(\alpha<2\).

	\subsection{Decoherence}
	
	\begin{figure*}[htbp]
		\centering
		\subfloat[Spatiotemporal magnitude \(|\psi(x,t)|\)]{\includegraphics[scale=0.3]{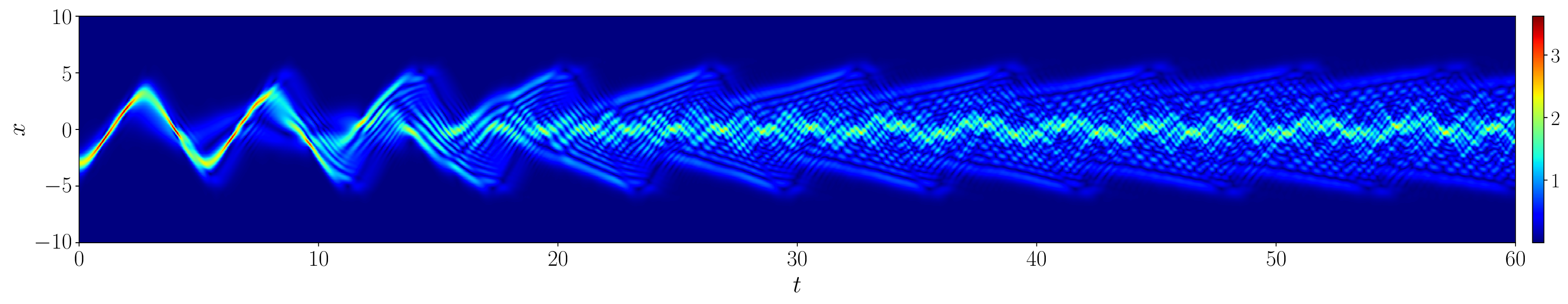}\label{subfig:focusing_alpha=1_5_mode=0_omega=-2_h=3_image_data}}\\
		\subfloat[Phase \(\arg \psi(x,t)\)]{\includegraphics[scale=0.3]{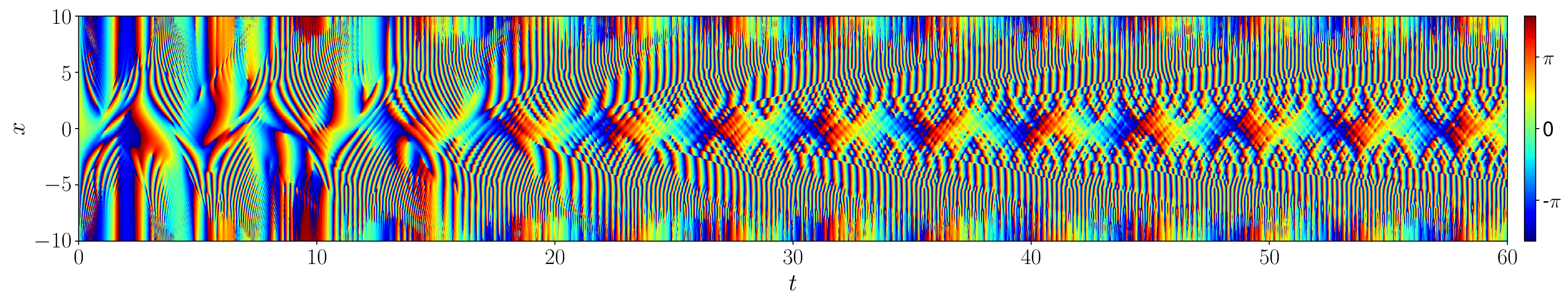}\label{subfig:focusing_alpha=1_5_mode=0_omega=-2_h=3_angle}}\\
		\subfloat[3D profile view]{\includegraphics[scale=0.55]{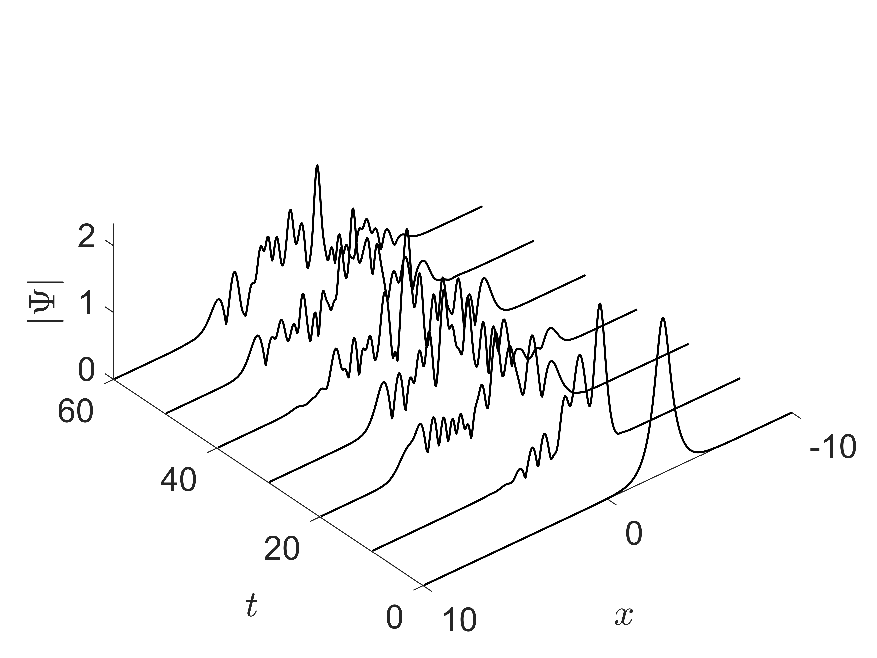}\label{subfig:focusing_alpha=1_5_mode=0_omega=-2_h=3_3D}}\quad
		\subfloat[Diagnostics \(X(t),S(t),M(t)\)]{\includegraphics[scale=0.55]{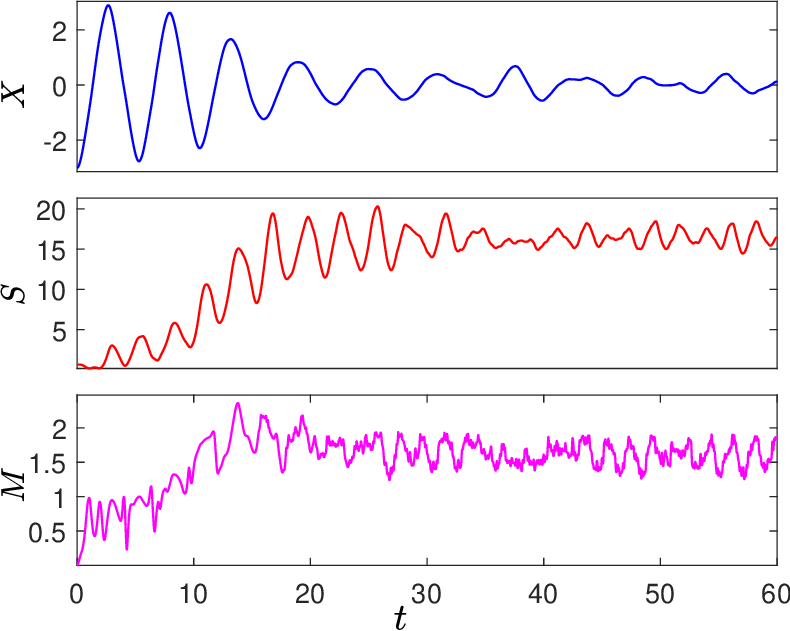}\label{subfig:focusing_alpha=1_5_mode=0_omega=-2_h=3_XSM}}
		\caption{Decoherent evolution in the focusing regime with \(\alpha=1.5\), \(n=0\), \(\Omega=-2\), and shift \(h=3\). Fractional focusing sharpens the initial profile but reduces stability: the magnitude breaks into peaks and troughs, the phase becomes irregular, and diagnostics show drift in \(X(t)\), growth in \(S(t)\), and a marked increase in \(M(t)\). These signatures indicate structural deformation. 
		}
		\label{fig:focusing_alpha=1_5_mode=0_omega=-2_h=3}
	\end{figure*}

	Figure~\ref{fig:focusing_alpha=1_5_mode=0_omega=-2_h=3} illustrates a representative decoherence scenario in the focusing regime for \(\alpha=1.5\), ground state (\(n=0\)), with frequency \(\Omega=-2\) and an initial shift \(h=3\). The spatiotemporal magnitude plot (Fig.~\ref{subfig:focusing_alpha=1_5_mode=0_omega=-2_h=3_image_data}) shows that the wave packet, initially localized, gradually loses coherence: localized structures develop peaks and troughs, indicating progressive destabilization. The phase evolution (Fig.~\ref{subfig:focusing_alpha=1_5_mode=0_omega=-2_h=3_angle}) highlights this breakdown, transitioning from smooth and regular patterns to increasingly irregular and chaotic oscillations. The 3D visualization (Fig.~\ref{subfig:focusing_alpha=1_5_mode=0_omega=-2_h=3_3D}) further confirms the structural instability, as the profile develops strong deformations over time.
	
	The diagnostics [Fig.~\ref{subfig:focusing_alpha=1_5_mode=0_omega=-2_h=3_XSM}] quantify the decoherence. The center of mass $X(t)$ evolves from sustained oscillations to an effectively stationary state, indicating the sloshing collapse. The variance $S(t)$ exhibits monotonic growth, consistent with progressive spreading and partial delocalization. The deformation measure $M(t)$ increases and approaches a nonzero plateau, signaling irreversible structural distortion and fragmentation. Collectively, these measures confirm that the wave packet undergoes breakdown.
	
	Compared with the classical Laplacian case \(\alpha=2\), the fractional model with \(\alpha=1.5\) amplifies instability: nonlocal interactions thin the core but make the state more fragile to perturbations. This behavior is consistent with the narrower stability windows observed in the spectral analysis for focusing nonlinearities. Thus, this example demonstrates how fractional dispersion destabilizes localized states in the focusing regime, leading to decoherence and structural breakdown.

	\section{Conclusions}\label{sec:concl}
	
	We have presented a systematic study of the fractional nonlinear Schr\"odinger (fNLS) equation in a harmonic trap, highlighting how replacing the Laplacian by its fractional power $\left(-\partial_x^2\right)^{\alpha/2}$, $\alpha\in(1,2]$, reshapes both spectra and dynamics. The novelty of this work is threefold: we employ a Fourier pseudo-spectral discretization with an explicit operator form (via a modified \texttt{FDMx}) suitable for Jacobian construction; we connect continuation of stationary branches with a full linearized spectral analysis; and we verify the spectral predictions through direct time integrations supported by quantitative diagnostics.
	
	The computations show that fractional dispersion of L\'evy type systematically shifts and bends the bifurcation curves $Q(\Omega)$, producing thinner, more sensitive focusing states and broader, less regular defocusing states as $\alpha$ decreases. Spectral analysis reveals that the stable intervals compress and fragment as the system moves away from the classical case $\alpha=2$, with excited modes losing stability earlier in the focusing regime, while defocusing branches maintain wider, though eventually broken, stability windows. The imaginary spectra further indicate that focusing and defocusing share closely related oscillatory bands—even when their growth rates differ—pointing to a common dynamical backbone shaped by nonlocal dispersion. Time-dependent simulations confirm these conclusions: near thresholds, we observe bounded, persistent oscillations. In the defocusing case, we find robust coherence and profile preservation. In contrast, in the focusing case, fractional dispersion promotes decoherence, drift, spreading, and profile deformation. The diagnostics $X(t)$, $S(t)$, and $M(t)$ provide a consistent, quantitative link from spectral stability to nonlinear behavior.
	
	Beyond these findings, the present study provides practical benchmarks for numerical implementations of fractional operators in confined nonlinear systems. Several directions for future research appear promising. The methodology employed here is primarily numerical, while a corresponding analytical treatment remains an open problem. Analytical results for the non-fractional counterpart near the linear limit have been presented previously in, e.g., \cite{zezyulin2012nonlinear, alfimov2019localized}. A detailed investigation of the effects of fractionality in the vicinity of the linear regime will be reported elsewhere. Extensions to higher dimensions could further elucidate the interplay between nonlocal dispersion, vortices, and solitons \cite{malomed2021optical, zhong2025fractional, lashkin2024three}. Constructing detailed stability maps for such higher-dimensional structures, similar to those reported here, also constitutes an important direction for future work.

	\section{Acknowledgements}
	\textbf{RK} acknowledges that this research is funded by the ITB Research Program 2026 under the ITB International Research Scheme through the Directorate of Research and Innovation, Institut Teknologi Bandung.
	\textbf{HS} acknowledges support by Khalifa University through  a Research \& Innovation Grant under project ID KU-INT-RIG-2024-8474000789.
	{In addition, the authors acknowledge the two anonymous referees for their constructive comments.}
	
	\section*{Data availability}
	No data was used for the research described in the article.
	
	
	\section*{Declaration of generative AI and AI-assisted technologies in the writing process}
	
	During the preparation of this work, the authors utilized Grammarly and ChatGPT to enhance language and readability. After using these tools/services, the authors reviewed and edited the content as needed and take full responsibility for the content of the publication.
	
	
	
	\bibliography{references}
	
\end{document}